\documentclass[twocolumn,prd,showpacs]{revtex4}
\usepackage{graphicx}

\begin{document}

\renewcommand{\theequation}{\thesection.\arabic{equation}}

\newcommand{\ud}{\mathrm{d}}

\newcommand{\hp}{\Omega_{hp0}}
\renewcommand{\sp}{\Omega_{sp0}}
\newcommand{\ph}{\Omega_{ph0}}
\renewcommand{\L}{\Omega_{\Lambda 0}}
\newcommand{\w}{\Omega_{w0}}
\renewcommand{\k}{\Omega_{K}}
\newcommand{\K}{\Omega_{K'0}}
\newcommand{\m}{\Omega_{m0}}
\renewcommand{\r}{\Omega_{r0}}
\newcommand{\g}{\Omega_{g0}}
\newcommand{\st}{\Omega_{st0}}

\newcommand{\be}{\begin{equation}}
\newcommand{\ee}{\end{equation}}
\newcommand{\bea}{\begin{eqnarray}}
\newcommand{\eea}{\end{eqnarray}}

\title{Phantom Friedmann Cosmologies and Higher-Order Characteristics of Expansion}

\author{Mariusz P. D\c{a}browski}
\email{mpdabfz@sus.univ.szczecin.pl}
\affiliation{\it Institute of Physics, University of Szczecin, Wielkopolska 15,
          70-451 Szczecin, Poland}
\author{Tomasz Stachowiak}
\email{toms@oa.uj.edu.pl}
\affiliation{\it Astronomical Observatory, Jagiellonian University, 30-244
Krak\'ow, ul. Orla 171, Poland}

\date{\today}

\input epsf

\begin{abstract}

We discuss a more general class of phantom ($p < -\varrho$) cosmologies with
various forms of both phantom ($w < -1$), and standard ($w > -1$) matter.
We show that many types of evolution which include both
Big-Bang and Big-Rip singularities are admitted and give explicit
examples. Among some interesting models, there exist non-singular oscillating
(or "bounce") cosmologies, which appear due to a competition between positive and negative
pressure of variety of matter content. From the point of view of the current observations
the most interesting cosmologies are the ones which start with a Big-Bang and terminate
at a Big-Rip. A related consequence of having a possibility of two types of
singularities is that there exists an unstable static universe approached
by the two asymptotic models - one of them reaches Big-Bang, and another reaches Big-Rip.
We also give explicit relations between density parameters
$\Omega$ and the dynamical characteristics for these generalized phantom models, including
higher-order observational characteristics such as jerk and "kerk".
Finally, we discuss the observational quantities such as luminosity distance, angular
diameter, and source counts, both in series expansion and
explicitly, for phantom models. Our series expansion formulas for the luminosity distance
and the apparent magnitude go as far as to the fourth-order in redshift $z$ term, which
includes explicitly not only the jerk, but also the "kerk" (or "snap") which may serve as
an indicator of the curvature of the universe.

\end{abstract}

\maketitle

\section{Introduction}

Phantom is matter which possesses a very strong negative pressure and so it violates
the null energy condition $\varrho + p>0$, where $\varrho$ is the energy density and $p$ is
the pressure. Consequently it violates all the remained energy conditions: the strong,
the weak and the dominant. In terms of the barotropic equation of state $p=w \varrho$
($w=$ const.) phantom is described as matter with $w < -1$. The
story of phantom and its observational support is not quite analogous
to the story of quintessence or dark energy - the matter which
violates the strong energy condition $\varrho+3p>0$, i.e., the matter which allows
$w<-1/3$. Namely, despite its theoretical investigations
\cite{striwall,AJIII}, and even admission for the early universe inflation, it was hardly
accepted as reality before the first supernovae data was revealed \cite{perlmutter}.
A different story happens with phantom.
A more detailed check of the data suggests that the current supernovae
not only gives evidence for quintessence with $-1<w<-1/3$, but also
admits phantom with $w<-1$ at a high confidence level
\cite{supernovae}. The point is that in view of the data from
supernovae there is {\it no sharp cut-off} of the admissible
values of $w$ at $w=-1$. Instead, the data even favours the values
of $w$ which are less than minus one.

Phantom matter was first investigated in the current cosmological
context by Caldwell \cite{caldwell99}, who
also suggested the name referring to its curiosity. In fact,
phantom (or ghost) must possess negative energy which leads to instabilities
on both classical and quantum level \cite{instabilities,nojiri04}.
Since it violates all possible energy conditions, it also puts in doubt
the pillars of relativity and cosmology such as: the positive mass theorems, the laws of black hole
thermodynamics, the cosmic censorship, and causality in relativity theory \cite{he,visser}.
In this context phantom becomes a real challenge for the theory, if
its support from the the supernovae data is really so firm.
The thing is that the main support for the phantom
comes from supernovae dimming rather than from large-scale structure and cosmic
microwave background observations. That is why some
alternative explanation of the dimming of supernovae
due to a conversion of cosmological term and photons into axions
was made \cite{kaloper}. From the theoretical point of view -
a release of the assumption of an analytic equation of state which
relates energy density and pressure and does not lead to energy conditions
violation (except for the dominant one) may also be useful
\cite{barrow04}. As for the explanation of the supernovae data
phantom is also useful in killing the doubled positive pressure
contribution in braneworld models \cite{braneIa}.

Phantom type of matter was also implicitly suggested in cosmological models with a
particle production \cite{john88}, in higher-order theories of gravity models
\cite{Pollock88}, Brans-Dicke models for $\omega < - \frac{3}{2}$, in non-minimally
coupled scalar field theories \cite{starob00}, in mirage cosmology of the
braneworld scenario \cite{kiritsis}, and in kinematically-driven quintessence (k-essence)
models \cite{chiba}, for example. The interest in phantom has
grown vastly during the last two years and various aspects of
phantom models have been investigated
\cite{Hann02,Frampt02,Frampt03,FramTaka02,McInnes,trodden1,
trodden2,Bastero01,Bastero02,abdalla04,Erickson,BIphantom,
LiHao03,singh03,nojiri031,nojiri032,nojiri033,nojiri034,simphan,onemli,diaz,zhang,cai}.

One of the most interesting features of phantom models is that they allow for a Big-Rip (BR)
curvature singularity, which appears as a result of having the infinite values of the scale
factor $a(t) \to \infty$ at a finite future.
However, as it was already mentioned, the evidence for phantom from observations is
mainly based on the assumption of the barotropic equation of state which tightly constraints
the energy density and the pressure. It is puzzling \cite{barrow04} that for Friedmann
cosmological models which do not admit an equation of state which links the energy density
$\varrho$ and the pressure $p$, a sudden future singularity of pressure may
appear. This is a singularity of pressure only, with finite energy
density which has an interesting analogy with singularities which appear in some inhomogeneous
models of the universe \cite{dabrowski93,dabrowski95,inhosfs}.

Recently, phantom cosmologies which lead to a quadratic polynomial in
canonical Friedmann equation have been investigated \cite{phantom1}. However,
only the two types of phantom matter with the equations of state $p=-(4/3)\varrho$
(phantom) and $p=-(5/3)\varrho$ (superphantom) were admitted.
In this paper we extend our discussion into four phantom fluids by adding
subphantom ($p=-2\varrho$), and hyperphantom ($p=-(7/3)\varrho$),
and consider all types of matter with $-7/3 \leq w \leq 1$ which completes
a general discussion of the Friedmann equation with many-fluid matter source given
in series of papers \cite{FRWelliptic}. As it was found in \cite{phantom1}, there exist
interesting dualities between phantom and ordinary matter models
which are similar to dualities in superstring cosmologies
\cite{meissner91,superjim}. These dualities were generalized
to non-flat and scalar field models \cite{lazkoz1,lazkoz2}, brane models
\cite{calcagni}, and found to also be
related to ekpyrotic models \cite{ekpyrotic,triality}.

In Section II we write down the field equations for Friedmann
cosmology which involve all standard and phantom matter fluids and
express them in terms of the observational parameters such as
the dimensionless energy density parameters $\Omega$,
Hubble parameter $H$, deceleration parameter $q$, and the higher
derivative parameters like jerk $j$, "kerk" $k$ etc. \cite{jerk,snap,weinberg} which
may serve as indicators of the equation of state (statefinders)
and the curvature of the universe. We also
give explicit expressions relating these quantities between
themselves, which allows to eliminate some of them, when the
comparison with astronomical data is required. Some interesting
symmetries in these relations, which refer to the appearance of the
density parameters, are very clearly visible.

In Section III we study some exact cosmological models which lead
to a polynomial of the third and the fourth order in the canonical
Friedmann equation. This is the next step to what has been done in
our earlier reference \cite{phantom1}, since now the solutions are
given in terms of the elliptic functions only. Because of the
tenth degree polynomial in the basic canonical equation which
involves all the fluids from $w=-7/3$ to $w=1$, we discuss only the
models which, after some change of variables, can be reduced to the
case of the third or fourth degree canonical equation polynomial.
All the possible cosmological models which are allowed in such a
general case are discussed in the two appendices A and B.

In Section IV we present the discussion of the observational
quantities (luminosity distance, angular diameter, source counts)
both in terms of the series expansion which involves all the
fluids under studies together with the higher-derivative parameters, and in
some cases explicitly given by elliptic functions. Similar results
were obtained for quintessence ($-1<w<0$) models in an archaic reference \cite{AJIII}.
In Section V we give the conclusions.

\section{Friedmann equation with phantom and observable parameters}
\setcounter{equation}{0}

Phantom is a new type of cosmological fluid of a very
strong negative pressure which violates the null energy condition
(NEC) \cite{he,visser}
\be
p + \varrho >0~,
\ee
i.e., it obeys a barotropic equation of state
\be
p = \left({\gamma - 1}\right) \varrho = w \varrho,
\ee
with {\it negative} barotropic index
\be
\gamma = w + 1 < 0~,
\ee
or, as more commonly expressed nowadays, with
\be
w < -1~.
\ee
Phantom also violates the strong, weak, and dominant energy
conditions (SEC, WEC, and DEC, respectively)
\bea
\varrho + p &>& 0, \hspace{1.cm}{\rm and} \hspace{1.cm}\varrho + 3p > 0~,\\
\varrho &>& 0, \hspace{1.cm}{\rm and} \hspace{1.cm}\varrho + p > 0~,\\
\varrho &>& 0, \hspace{1.cm} {\rm and} \hspace{1.cm}- \varrho < p < \varrho~.
\eea

Phantom may be included into the basic
system of equations for isotropic and homogeneous Friedmann universe which reads as
\bea
\label{perho1}
\kappa^2 \varrho & = & - \Lambda + 3 \frac{K}{a^2} + 3
\frac{\dot{a}^2}{a^2} ,\\
\label{perho2}
\kappa^2 p & = & \Lambda - 2 \frac{\ddot{a}}{a} - \frac{K}{a^2} -
\frac{\dot{a}^2}{a^2}  ,
\eea
where $a(t)$ is the scale factor, $K = 0, \pm 1$ the curvature
index, $\Lambda$ the cosmological constant, $\kappa^2 = 8\pi G$ -- the Einstein constant.

Using equations (\ref{perho1}) and (\ref{perho2}), one gets the Friedmann equation in the
form
\begin{equation}
\label{Friedroro}
\frac{\dot{a}^2}{a^2} = \frac{\kappa^{2}}{3} \rho - \frac{K}{a^{2}} +
\frac{\Lambda}{3}.
\end{equation}
After imposition of the conservation law
\be
\label{conslaw}
\varrho a^{3\gamma} = (3/\kappa^2) C_{\gamma} =
{\rm const.}~,
\ee
one gets Eq. (\ref{Friedroro}) as follows
\begin{equation}
\label{FriedCCC}
\frac{1}{a^2} \left( \frac{da}{dt} \right)^2 =
\frac{C_{\gamma}}{a^{3\gamma}}  -
\frac{K}{a^2} + \frac{\Lambda}{3} .
\end{equation}
The standard types of cosmological matter are: stiff-fluid ($\gamma = 2$),
cosmic gas ($\gamma = 5/3$),
radiation ($\gamma = 4/3$), dust ($\gamma = 1$), cosmic strings ($\gamma = 2/3$),
domain walls ($\gamma = 1/3$), and cosmological constant ($\gamma = 0$) and the models
which involve them have been studied in detail \cite{FRWelliptic,AJIII}. They, in fact,
admit a large class of oscillating (non-singular) solutions due to
a balance between the positive and negative pressure domination
periods. This is in full analogy to what happens in ekpyrotic
scenario \cite{ekpyrotic}, where the positive
potential energy dominates throughout some period of the evolution
(including now), and the negative potential energy dominates near
the bounce.

In this paper, which follows our earlier Ref. \cite{phantom1}, we extend the discussion
of the models onto the case of {\it negative} barotropic index phantom matter with
$\gamma = -1/3$ (phantom), $\gamma = -2/3$ (superphantom), $\gamma = -1$ (subphantom),
and $\gamma = -4/3$ (hyperphantom). Due to a phantom duality
\cite{phantom1,lazkoz1,lazkoz2,triality}
\be
\gamma \to -\gamma~,
\ee
these phantom types of matter are dual to domain walls ($\gamma=1/3$), cosmic strings
($\gamma=2/3$), dust ($\gamma=1$) and radiation ($\gamma=4/3$), respectively.

The simplest way to consider these dualities is to integrate the
Friedmann equation (\ref{FriedCCC}) for one general fluid only,
which gives
\be
\label{aforstan}
a(t) = a_0 \mid t \mid^{\frac{2}{3\gamma}}~,
\ee
for ordinary ($\gamma>0$) matter, and
\be
\label{aforphan}
a(t) = a_0 \mid t \mid^{-\frac{2}{3\mid \gamma \mid}}~,
\ee
for phantom $\gamma = - \mid \gamma \mid <0$. In both cases there is a curvature singularity
at $t=0$, but in the former case it is of a Big-Bang type, while in the latter case it is of
a Big-Rip type. From the current observations, however, it is reasonable
to choose the solution (\ref{aforstan}) for positive times $t>0$,
and the solution (\ref{aforphan}) for negative times $t<0$. In
this context the age of the universe in a standard matter case is
the time from Big-Bang at $t=0$, to the present time $t=t_0$, and reads as
\be
\label{agestandard}
t_0 = \frac{2}{3\gamma} \frac{1}{H_0}~,
\ee
while in a phantom case one can only calculate the time from the
present $t=-t_0$ to a Big-Rip at $t=0$ which reads as
\be
\label{agephantom}
t_0 = \frac{2}{3\mid \gamma \mid} \frac{1}{H_0}~.
\ee
Similar investigations should be taken into account, if one wants
to calculate the redshift of an observed galaxy. If a light ray
was emitted at the time $t_1$ in the past, then in the standard matter
case, the redshift reads as
\be
\label{z}
1+z = \frac{a(t_0)}{a(t_1)} = \left( \frac{t_0}{t_1}
\right)^{\frac{2}{3\gamma}}~.
\ee
We can see that the redshift tends to infinity, if $t_1$ approaches
zero (i.e. at Big-Bang). The situation is different for phantom, since
the redshift is given by
\be
1+z = \frac{a(t_0)}{a(t_1)} = \left( \frac{t_1}{t_0}
\right)^{\frac{2}{3\mid \gamma \mid}}~,
\ee
and the redshift tends to infinity if the time $t_1$ approaches
minus infinity. On the other hand, if the time of emission of the
light ray is the time of Big-Rip, which is at $t_1=0$, then we have
formally $z=-1$.

Let us now define the appropriate cosmological parameters which characterize the dynamics of the
evolution of the universe which are:
\\
\noindent the Hubble parameter
\be
\label{hubb}
H = \frac{\dot{a}}{a}~,
\ee
the deceleration parameter
\be
\label{dec}
q  =  - \frac{1}{H^2} \frac{\ddot{a}}{a} = - \frac{\ddot{a}a}{\dot{a}^2}~,
\ee
the jerk parameter \cite{jerk}
\be
\label{jerk}
j = \frac{1}{H^3} \frac{\dddot{a}}{a} =
\frac{\dddot{a}a^2}{\dot{a}^3}~,
\ee
and the "kerk" (snap \cite{snap}) parameter (which includes the fourth derivative of the
scale factor $a$ - we can carry on to call the higher derivative parameters
"lerk", "merk", "nerk", "oerk" etc., respectively)
\be
\label{kerk}
k = -\frac{1}{H^4} \frac{\ddot{\ddot{a}}}{a} =
-\frac{\ddot{\ddot{a}}a^3}{\dot{a}^4}~.
\ee

The application of the definitions of the parameters (\ref{hubb}), (\ref{dec}),
{\ref{jerk}), and (\ref{kerk}) gives the following equalities
\bea
\dot{H} &=& -H^2 \left( q+1 \right)~,\\
\ddot{H} &=& H^3 \left( j+3q+2 \right)~,\\
\dddot{H} &=& -H^4 \left[ k+4j+3q \left( q+4 \right)+6 \right]~.
\eea

A comparison of phantom models with observational data
requires the introduction of dimensionless density parameters \cite{AJIII,braneIa}
\bea
\label{Omegadef}
\Omega_{x0}  &=&  \frac{\kappa^2}{3H_0^2} \varrho_{x0} ,\\
\Omega_{K0}  &=&  \frac{K}{H_0^2a_0^2} ,\\
\Omega_{\Lambda_0}  &=&  \frac{\Lambda_0}{3H_0^2} ,
\eea
where $x \equiv hp, bp, sp, ph, w, s, m, r, g, st$ (hyperphantom, subphantom,
superphantom, phantom, domain walls, cosmic strings,
dust, radiation, cosmic gas, and stiff-fluid, respectively),
and the index "0" means that we take the parameters at the present
moment of the evolution $t=t_0$.

Applying the relation ($K'=C_s-K$, where $C_s=C_{\gamma}$ for
$\gamma=2/3$, cf. (\ref{conslaw}))
\be
\label{kaprime}
\Omega_{K'0} = \Omega_{s0} - \Omega_{K0} ,
\ee
we realize that for $a=a_0$, the Friedmann equation (\ref{FriedCCC}) can be written down
in the form
\bea
\label{Om=1}
&&\Omega_{hp0} + \Omega_{bp0} + \Omega_{sp0} + \Omega_{ph0} +
\Omega_{\Lambda_0}  \\
&& + \Omega_{w0} + \Omega_{K'0}
+ \Omega_{m0} + \Omega_{r0} + \Omega_{g0} + \Omega_{st0} = 1  .\nonumber
\eea

From the field equations (\ref{perho1}) and (\ref{perho2}) we can also derive
the following relations between the observational parameters
\bea
\label{Ka}
\frac{K}{H_0^2a_0^2} &=& 3 \Omega_{st0} + \frac{5}{2} \Omega_{g0} + 2 \Omega_{r0} +
\frac{3}{2} \Omega_{m0} \nonumber \\
&+& \Omega_{s0} + \frac{1}{2} \Omega_{w0}
- \frac{1}{2} \Omega_{ph0}  \\
&-& \Omega_{sp0} - \frac{3}{2} \Omega_{bp0} -
2 \Omega_{hp0} - q_0 - 1~,\nonumber
\eea
\bea
\label{Kaprim}
-\frac{K'}{H_0^2a_0^2} &=& 3 \Omega_{st0} + \frac{5}{2} \Omega_{g0} + 2 \Omega_{r0} +
\frac{3}{2} \Omega_{m0} \nonumber \\
&+& \frac{1}{2} \Omega_{w0}
- \frac{1}{2} \Omega_{ph0}  \\
&-& \Omega_{sp0} - \frac{3}{2} \Omega_{bp0} -
2 \Omega_{hp0} - q_0 - 1~, \nonumber
\eea
and
\bea
\label{Lambda0}
\frac{\Lambda_0}{3H_0^2} &=& 2 \Omega_{st0} + \frac{3}{2} \Omega_{g0} + \Omega_{r0} + \frac{1}{2} \Omega_{m0} -
q_0 - \frac{1}{2} \Omega_{w0} \nonumber \\
&-& \frac{3}{2} \Omega_{ph0} - 2
\Omega_{sp0} - \frac{5}{2} \Omega_{bp0} - 3
\Omega_{hp0}~.
\eea
Alternatively, one can use (\ref{Ka}) and (\ref{Lambda0}) to express the deceleration
parameter $q_0$ as
\bea
\label{q0elim}
q_0 &=& 2 \Omega_{st0} + \frac{3}{2} \Omega_{g0} + \Omega_{r0} + \frac{1}{2} \Omega_{m0}
 - \frac{1}{2} \Omega_{w0}  \\
&-& \Omega_{\Lambda_0} - \frac{3}{2} \Omega_{ph0} - 2
\Omega_{sp0} - \frac{5}{2} \Omega_{bp0} - 3
\Omega_{hp0}~. \nonumber
\eea
Note that cosmic string energy density parameter $\Omega_{s0}$
does not appear in (\ref{q0elim}). Similarly, it is possible to
express the jerk parameter
\bea
\label{j0elim}
j_0 &=& 10 \Omega_{st0} + 6 \Omega_{g0} + 3\Omega_{r0} + \Omega_{m0}
+ \Omega_{\Lambda_0} \nonumber \\
&+& 3 \Omega_{ph0} + 6
\Omega_{sp0} +10 \Omega_{bp0} +15 \Omega_{hp0}~.
\eea
In fact, neither cosmic string energy density parameter
$\Omega_{s0}$, nor domain wall energy density parameter $\Omega_{w0}$, do not
appear in (\ref{j0elim}). Next step is to express the kerk
parameter by
\bea
\label{k0elim}
k_0 &=& q_0 j_0 + 60 \Omega_{st0} + 30 \Omega_{g0} + 12 \Omega_{r0} + 3\Omega_{m0}
 \nonumber \\
&-& 3 \Omega_{ph0} - 12
\Omega_{sp0} - 30 \Omega_{bp0} - 60 \Omega_{hp0}~.
\eea
Here, in turn, none of the three parameters: cosmic string energy density parameter
$\Omega_{s0}$, domain wall energy density parameter $\Omega_{w0}$, and  the cosmological
term $\Omega_{\Lambda_0}$ does not appear explicitly in (\ref{k0elim}), although they are
included in both $q_0$ and $j_0$.

\section{Examples of phantom Friedmann cosmologies}
\setcounter{equation}{0}

Using the new variables
\be
\label{param}
y = \frac{a}{a_0}, \hspace{0.5cm} \tau = H_0 t ,
\ee
one turns the basic Friedmann equation (\ref{FriedCCC}) into the form
\bea
\label{FRWbasic}
&&\left( \frac{dy}{d\tau} \right)^2 = \Omega_{hp0}y^6 + \Omega_{bp0}y^5 + \Omega_{sp0}y^4
\nonumber \\
&& + \Omega_{ph0}y^3 + \Omega_{\Lambda_0}y^2 +
\Omega_{w0} y + \Omega_{K'0}  \\
&& + \Omega_{m0} y^{-1} + \Omega_{r0} y^{-2} + \Omega_{g0} y^{-3} +
\Omega_{st0} y^{-4}.\nonumber
\eea

The discussion now refers to an investigation of the cosmological
models which are implied by this Friedmann equation (\ref{FRWbasic}). Of course,
the integration of a general case would require the application of
abelian functions, since we have the tenth degree polynomial
to the right-hand side of (\ref{FRWbasic}). In this
paper we will discuss only the cases in which Eq. (\ref{FRWbasic})
can be reduced to the equation with the third or fourth degree
polynomial in the canonical equation. Phantom cosmologies which
result from the second degree polynomial were discussed in \cite{phantom1}.
Since a general mathematical discussion of the
canonical equations is quite extensive, we put most of it into the
Appendices A and B, leaving only some examples of solutions in the
main body of the paper. One example is out of the category of the
third degree polynomial in the canonical equation, and another one
is out of the category of the fourth degree polynomial. The last
example is out of the category of the second degree polynomial
(cf. \cite{phantom1}).

\subsection{Superphantom, $\Lambda$-term, strings, and radiation models}

The simplest case which can be integrable in terms of elliptic
functions is when we neglect all the fluids except for superphantom,
$\Lambda$-term, strings, and radiation in (\ref{FRWbasic}), i.e.,
\be
\label{sslam}
\left(\frac{\ud y}{\ud \tau}\right)^2 = \sp y^4+\L y^2+\K+\r y^{-2}~.
\ee
Since we deal with even powers of $y$ only, we can introduce a new variable
\be
x \equiv y^2~,
\ee
so that the equation (\ref{sslam}) reads as
\be
\left(\frac{dx}{2\tau}\right)^2 = \sp x^3+\L x^2+\K x+\r = W(x)~,
\ee
and this is an example of the third degree polynomial $W(x)$
in canonical equation discussed in the Appendix A.
From the definition (\ref{Omegadef}) we conclude that $\sp$ is always
greater than zero, and so we deal with the appropriate solutions of section A.1.
of Appendix A. The discriminant $\Delta$ of $W(x)$ is, in turn,
a polynomial of the second order in $\sp$
with a negative coefficient in front of $\sp^2$. Its roots are
\begin{equation}
        \Omega_{sp\pm} = \frac{18\r\K\L-4\K^3\pm(\K^2-3\r\L)^{3/2}}{54\r^2}
\end{equation}
The sign of $\Delta$ depends on the sign of $\K^2-3\r\L$.

For $\Delta$, in order to be positive, we must have
\be
\K^2>3\r\L, \hspace{0.5cm} {\mathrm and} \hspace{0.5cm}
\Omega_{sp-}<\sp<\Omega_{sp+}~.
\ee
The existence
of the appropriate values of the parameters follows from a general analysis of the
behaviour of the polynomial $W(x)$. For $\Delta>0$, the solutions
are given by (\ref{xp1a}) and (\ref{xm1a}), i.e.,
\bea
        y_-^2 &=& \frac{4}{\sp}\wp(2\tau-\tau_0+\omega_3)-\frac{\L}{3\sp} \nonumber\\
        y_+^2 &=& \frac{4}{\sp}\wp(2\tau-\tau_0)-\frac{\L}{3\sp} \nonumber,
\eea
with the invariants
\bea
    g_2 &=& \frac{1}{12}\L^2-\frac14\sp\K, \nonumber\\
    g_3 &=& \frac{1}{48}\sp\L\K-\frac{1}{216}\L^3-\frac{1}{16}\sp^2\r \nonumber.
\eea
The solution $y_-$ is oscillating and non-singular (ONS) type, while the
solution $y_+$ is singular and evolves from a Big-Rip to a Big-Rip (BR-BR type)
(compare Fig. \ref{fign311}).

Because of the change of the variable, $2\tau$ appears as an argument, and the "physical"
period is not $2\omega_1$, but $\omega_1$. For the solution $y_-$ to be possible,
the roots of $W$ must be such that $e_3 < 1 < e_2$. This condition can be simplified
by analyzing the position of the minimum of $W$. It yields
\bea
\L < \frac12\K-\frac32\sp~.\nonumber
\eea
When the opposite is true, the other solution is valid. Moreover, as $W(0)=\r\geq 0$,
we have $e_3 \leq 0$, so that the oscillating solution $y_+$ necessarily
possesses the $y = 0$ singularity. For the $y_-$ solution, we obtain the singular
behaviour when $\sp\K > 0$ and $\L > 0$. Otherwise $e_1 > 0$.

If $\Delta$ is to be equal to zero (case A.1.b. of Appendix A), its discriminant must
be non-negative, i.e.,
\be
\K^2 \geq 3\r\L, \hspace{0.5cm} \textrm{ and} \hspace{0.5cm}
\sp=\Omega_{sp\pm}~.
\ee
On the other hand, the root of $W(x)$ is triple, if
\bea
        &&\Delta = \L^2-3\sp\K=0, \\
\noindent \textrm{ and}\nonumber\\
        &&9\sp\L\K-2\L^3-27\sp^3\r=0~, \nonumber \label{triple_cond}
\eea
so that the solutions are
\bea
\widetilde{y}_0^2 &=& -\sqrt[3]{\frac{\widetilde{\r}}{\widetilde{\sp}}},
\nonumber\\
\noindent \textrm{ and} \nonumber \\
y_0^2 &=& -\sqrt[3]{\frac{\r}{\sp}}+\frac{4}{\sp(2\tau-\tau_0)^2}.
\eea
Here the tilde indicates, that the Hubble constant $H_0=0$, so that the normalization
is not possible. As the root is non-positive ($W(0) \geq 0$),
the solution $y_0$ is both BB and BR singular, whereas the solution
$\widetilde{y}_0$ is not physical (compare Fig. \ref{x01b}).

Another possibility is a double root. Both of the conditions (\ref{triple_cond}),
must be broken then (if only one was, the discriminant would not be equal to
zero). There are two subcases, depending on the sign of $g_3$.

With $g_3>0$, we have a stable static (SS) Einstein universe $y=0$, if $\K=\r=0$, and another
model of the universe given by
\begin{equation}
    y_0^2 = \frac{2\sqrt[3]{g_3}}{\sp}
    \left\{2+3\tan^2\left[\sqrt{\frac{3}{2}}\sqrt[6]{g_3}(\tau-\tau_0)\right]\right\}
        -\frac{\L}{3\sp},
\end{equation}
which can be of BR-BR type, if $\Omega_{\Lambda 0}<0$, and BB-BR
type, if $\Omega_{\Lambda 0}>0$.

For $g_3<0$ the solutions are exactly the same as in section A.1.b. of Appendix
A, i.e.,
\bea
    y_-^2 &=& \frac{2\sqrt[3]{g_3}}{\sp}\left\{2-3\tanh^2\left[\sqrt{\frac{3}{2}}\sqrt[6]{-g_3}(\tau-\tau_0)\right]\right\}
        -\frac{\L}{3\sp},\nonumber \\
    y_+^2 &=& \frac{2\sqrt[3]{g_3}}{\sp}\left\{2-3\coth^2\left[\sqrt{\frac{3}{2}}\sqrt[6]{-g_3}(\tau-\tau_0)\right]\right\}
        -\frac{\L}{3\sp}, \nonumber \\
    y_0^2 &=& -\frac{2\sqrt[3]{g_3}}{\sp}-\frac{\L}{3\sp}\textrm{ (static).}
\eea
The first solution is non-singular, if a static solution $y=0$ exists.
The second is BB singular, and the third one might be free from the
singularity, if the double root $0 < y_0 <1$.

Finally, $\Delta$ is negative, when $\sp < \Omega_{sp-}$ or $\sp > \Omega_{sp+}$. For
$\K^2 < 3\r\L$, this is the case for all values of $\sp$. The solution is that of
Appendix A.1.c. and, as above, $W(0)\geq 0$, which means it is
not only BR singular, but also BB singular, despite the case presented in Fig. \ref{n313}.
The exact solution is
\begin{equation}
        y_0^2 = \frac{4}{\sp}\wp(2\tau-\tau_0)-\frac{\L}{3\sp}~,
\end{equation}
with ``period'' being $2{\mathcal Re}(\omega_1)$.

\subsection{Hyperphantom, $\Lambda$-term, and radiation models}

The Firedmann equation (\ref{FRWbasic}) now reduces to
\begin{equation}
\label{hplamr}
    \left(\frac{\ud y}{\ud u}\right)^2 = \hp y^6 + \L y^2 + \r y^{-2}.
\end{equation}
We use the same variables as in the superphantom case
\begin{equation}
    x\equiv y^2,\qquad \tau\equiv u,
\end{equation}
so that (\ref{hplamr}) reduces to
\begin{equation}
    \left(\frac{\ud x}{2\ud\tau}\right)^2 = \hp x^4 + \L x^2 + \r, \label{ref1}
\end{equation}
which can be easily analyzed, as it is biquadratic. This is an example of the
fourth degree polynomial $W(x)$ in the canonical equation discussed in the Appendix B.
The four roots are given by
\begin{equation}
        x_{0\pm}^2 = \frac{-\L\pm\sqrt{\L^2-4\hp\r}}{2\hp},
\end{equation}
where $\L=1-\hp-\r$, and we can analyze the subcases with respect to the
$(\hp,\r)$ parameter space, since $\L$ can both be negative and positive, so
it always exists for given $\hp$ and $\r$.

The expression under the root is the discriminant $\Delta_2$ of the
polynomial in equation (\ref{ref1}), when considered as quadratic in
$x^2$, i.e.,
\begin{equation}
        \Delta_2 = \hp^2 + \r(\r -2) - 2\hp(\r +1) + 1.
\end{equation}
All signs of $\Delta_2$ are possible for positive values of the
parameters, and we consider them consecutively.

\subsubsection{$\Delta_2<0$}

In this case the roots $x_{0\pm}^2$ are complex and different, so there are four distinct
complex roots of the main polynomial (\ref{hplamr}). Accordingly, this is the case B.1.a.1 of
the general classification from Appendix B. The evolution of the model is very interesting from the current
observational point of view, since it starts from Big-Bang and
terminates at Big-Rip (compare Fig. \ref{n4111}).

\subsubsection{$\Delta_2>0$}

Here the roots $x_{0\pm}^2$ are real, so we analyse their signs, which, in turn,
depend on the sign of $\L=1-\hp-\r$. It is worth noticing that for both these
cases we always have
\begin{equation}
        |\L|>\sqrt{\Delta_2}. \label{ref2}
\end{equation}

For $\L>0$ we have $\r<1-\hp$, and the roots
$x_{0\pm}^2$ are negative thanks to the inequality (\ref{ref2}), and this
is the case B.1.a.1 of Appendix B again.

For $\L<0$, we have $\r>1-\hp$, and now both roots $x_{0\pm}^2$ are positive, so the main polynomial has four
distinct real roots. This corresponds to the general case B.1.a.3.
As there are two positive and two negative roots, a possible oscillating
solution is singular at $y=0$ (BB-BR type). There is also a
bouncing Big-Rip solution only (cf. Fig. \ref{n4113}).

\subsubsection{$\Delta_2=0$}

The conditions for the parameters are
\begin{equation}
        \hp = (1\pm\sqrt{\r})^2 \Leftrightarrow \r = (1\pm\sqrt{\hp})^2,
\end{equation}
which imply that
\be
\L = -2\sqrt{\r}(\sqrt{\r}\pm1) = -2\sqrt{\hp}(\sqrt{\hp}\pm1)~.
\ee
We thus obtain two double roots, given by
\begin{equation}
        x_{0\pm} = -\frac{\L}{2\hp}=\frac{\sqrt{\hp}\pm1}{\sqrt{\hp}}.
\end{equation}

If $\L<0$ we have two double real roots - one pair negative, one positive.
That is, the general case B.1.c.1 so that there are two unstable
static (US) solutions. Besides, there is an interesting asymptotic
solution which is monotonic from one static solution to the other
(MUS-MUS type), and the two bouncing solutions - one of them
asymptotes from a static towards Big-Bang and another from a static towards
Big-Rip (cf. Fig. \ref{n4131}).

If $\L>0$ we have two double complex roots, which is the general case B.1.c.2
- the solution is of BB-BR type (cf. Fig. \ref{n4132}).

Finally, for $\L=0$ we obtain a quadruple root $x=0$, and the general
solution is that of B.1.e. However, since an unstable static universe
is at $y=0$, then only one of the asymptotic solutions is physical and is again
a BB-BR type (cf. Fig. \ref{n415}).

\subsection{Superphantom and radiation models}

Allowing superphantom ($\gamma=-2/3$) and radiation ($\gamma = 4/3$) only in the universe
we get from (\ref{param}) and (\ref{FRWbasic}) (cf. Ref.
\cite{phantom1})
\bea
\label{suprad}
\left( \frac{dy}{d\tau} \right)^2 = \Omega_{sp0}y^4 +
\Omega_{r0} y^{-2}~.
\eea
The equation (\ref{suprad}) solves parametrically by
\bea
a(\eta) & =& \frac{a_0}{\Omega_{sp0}^{\frac{1}{6}}}
\left[\frac{1}{4} \Omega_{sp0}^{2} (\eta - \eta_0)^2 -
\Omega_{r0} \right]^{\frac{1}{6}}~,\\
t(\eta) & =& \frac{\Omega_{sp0}^{\frac{2}{3}}}{H_0}
\int{\frac{d\eta}{\left[\frac{1}{4} \Omega_{sp0}^{2} (\eta - \eta_0)^2 -
\Omega_{r0} \right]^{\frac{2}{3}}}}~.
\eea
The solution is a BB-BR type. The Big-Bang
sigularity at $ a \to 0$ appears for
\be
\eta_{BB} = \eta_0 + \frac{2\sqrt{\Omega_{r0}}}{\Omega_{sp0}}~,
\ee
while
the Big-Rip singularity at $a \to \infty$ appears for
\be
\eta - \eta_0 \to \infty~.
\ee
The energy density of radiation and superphantom equality time
is given by
\be
\eta_{eq} = \frac{2 \eta_0}{\Omega_{sp0}}
\sqrt{\frac{a_0^6}{2\Omega_{sp0}} + \Omega_{r0}}~.
\ee
This is the best exact solution to investigate the properties of
phantom model evolution from Big-Bang to Big-Rip, since it
integrates in elementary functions. As it was shown in Ref.
\cite{phantom1} this is not the case for dust ($\gamma =0$) and phantom
($\gamma = -2/3)$ only model.

\section{Observational quantities for general phantom models}
\setcounter{equation}{0}

\subsection{$O(z^4)$ luminosity distance $D_{L}(z)$ formula}

We will follow standard derivation of the luminosity distance
\cite{weinberg}, but include higher order characteristics such as jerk
\cite{jerk} and "kerk" (\ref{kerk}) as state-finders. The physical
distance $D$ which is travelled by a light ray emitted at the time
$t_1$, and received at the time $t_0$ is given by
\be
\label{D}
D = c \int_{t_1}^{t_0} dt = c (t_0 - t_1)~.
\ee
The scale factor $a(t)$ and its inverse $1/a(t)$ at any moment of time $t$ can
be obtained as series expansion around $t_0$ as ($a(t_0) \equiv a_0$)
\bea
\label{seriesa}
&&a(t) = a_0 \left\{ 1 + H_0 (t-t_0) - \frac{1}{2!}q_0 H_0^2
(t-t_0)^2 \right.  \\
&& \left. + \frac{1}{3!} j_0 H_0^3 (t-t_0)^3 - \frac{1}{4!} k_0
H_0^4 (t-t_0)^4 + O[(t-t_0)^5]\right\}~, \nonumber
\eea
and
\bea
\label{seriesinva}
&&\frac{1}{a(t)} = \frac{1}{a_0} \left\{ 1 + H_0 (t_0-t) + H_0^2
\left(\frac{q_0}{2} +1 \right) (t_0-t)^2 \right. \nonumber \\
&& \left. + H_0^3 \left(q_0 + \frac{j_0}{3} + 1 \right) (t_0-t)^3
\right.  \\
&& \left. + H_0^4 \left(1 + \frac{j_0}{3} + \frac{q_0^2}{4} + \frac{3}{2} q_0
+ \frac{k_0}{24} \right) (t_0-t)^4 \right. \nonumber \\
&& \left. + O[(t_0-t)^5]\right\}~, \nonumber
\eea
Putting $t=t_1$ in (\ref{seriesinva}), and using (\ref{z}) and
(\ref{D}), we get a series $z=z(D)$ as follows
\bea
&& 1+z = 1 + \left( \frac{H_0 D}{c} \right) + \frac{1}{2} \left(q_0 +
2 \right) \left( \frac{H_0 D}{c} \right)^2 \nonumber \\
&&+ \frac{1}{6} \left[j_0 + 6(q_0 + 1) \right] \left( \frac{H_0 D}{c}
\right)^3 \nonumber \\
&& + \frac{1}{24} \left[k_0 + 8 j_0 + 6q_0(q_0 + 6) + 24
\right] \left( \frac{H_0 D}{c} \right)^4 \nonumber \\
&& + O \left[\left( \frac{H_0 D}{c}
\right)^5 \right]~.
\eea
This series can be inverted to get $D=D(z)$, i.e.,
\bea
\label{D(z)}
&& D(z) = \frac{cz}{H_0} \left\{ 1 - \left(1 + \frac{q_0}{2}
\right)z + \left(1 + q_0 + \frac{q_0^2}{2} - \frac{j_0}{6} \right) z^2
\right. \nonumber \\
&& \left. + \left[ \frac{j_0}{2} \left(\frac{5}{6} q_0 +1 \right) -
\frac{k_0}{24} - \frac{5}{8} q_0^3 - \frac{3}{2} q_0 (q_0 + 1) - 1
\right] z^3 \right. \nonumber \\
&& \left. + O(z^4) \right\}~.
\eea
The luminosity distance $D_L$ is defined as the distance to an
object whose energy flux falls off in the way, as if we dealt with
a Euclidean space \cite{weinberg}, i.e.,
\be
F = \frac{L}{4\pi D_L^2}~,
\ee
where $L$ is the total brightness of the source. In the Friedmann universe
it reads as
\begin{equation}
D_L = (1+z)a_0 r_0 = (1+z)a_0 S_K(\chi)  , \label{lum}
\end{equation}
where $r_0$ is the radial distance from a source to an observer, and
\bea
S_K(\chi) = \left\{
        \begin{array}{l}
            \frac{1}{\sqrt{K}}\sin(\sqrt{K}\chi),  K>0\\
            \chi,  K = 0\\
            \frac{1}{\sqrt{|K|}}\sinh(\sqrt{|K|}\chi),  K<0
        \end{array}\right. \
\eea
From the null geodesic equation in the Friedmann universe we have
\be
\label{geodeq}
\int_{t_1}^{t_0} \frac{c dt}{a(t)} = \int_{0}^{r_0}
\frac{dr}{\sqrt{1-Kr^2}} = \chi(r_0) = S_K^{-1}(r_0)~,
\ee
so that we can conclude that
\be
r_0 = S_K(\chi) = S_K \left( \int_{t_1}^{t_0} \frac{c dt}{a(t)}
\right)~,
\ee
which can be expanded in series for small distances as
\be
\label{r0ser}
r_0 = \left( \int_{t_1}^{t_0} \frac{c dt}{a(t)}
\right) - \frac{K}{3!} \left( \int_{t_1}^{t_0} \frac{c dt}{a(t)}
\right)^3 + O \left[ \left( \int_{t_1}^{t_0} \frac{c dt}{a(t)}
\right)^5 \right]~.
\ee
Now, the problem reduces to a calculation of an integral
(\ref{geodeq}) in terms of the distance $D$. For this sake, we use
series expansion (\ref{seriesinva}), i.e.,
\bea
&&\int_{t_1}^{t_0} \frac{c dt}{a(t)} =
\nonumber \\
&& = \frac{c}{a_0} \int_{t_1}^{t_0} dt
\left\{ 1 + H_0 (t_0-t) + H_0^2
\left(\frac{q_0}{2} +1 \right) (t_0-t)^2 \right. \nonumber \\
&& \left. + H_0^3 \left(q_0 + \frac{j_0}{3} + 1 \right) (t_0-t)^3
\right.  \nonumber \\
&& \left. + H_0^4 \left(1 + \frac{j_0}{3} + \frac{q_0^2}{4} + \frac{3}{2} q_0
+ \frac{k_0}{24} \right) (t_0-t)^4
\right. \nonumber \\
&& \left. + O[(t_0-t)^5]\right\}~,
\eea
which integrates to give
\bea
\label{r01}
\int_{t_1}^{t_0} \frac{c dt}{a(t)} &=& \frac{D}{a_0}
\left\{ 1 + \frac{1}{2} \left( \frac{H_0 D}{c} \right) \right. \nonumber \\
 &+& \left. \frac{1}{3} \left(\frac{q_0}{2} + 1 \right) \left( \frac{H_0 D}{c} \right)^2
\right. \nonumber \\
&+& \left. \frac{1}{4} \left(q_0 + \frac{j_0}{3} + 1 \right) \left( \frac{H_0 D}{c}
\right)^3 \right.  \nonumber \\
&+& \left. \frac{1}{5} \left(1 + \frac{j_0}{3} + \frac{q_0^2}{4} + \frac{3}{2} q_0
+ \frac{k_0}{24} \right) \left( \frac{H_0 D}{c} \right)^4
\right. \nonumber \\
&+& \left. O[\left( \frac{H_0 D}{c} \right)^5]\right\}~.
\eea
Notice that
\bea
\label{r03}
&& \left( \int_{t_1}^{t_0} \frac{c dt}{a(t)} \right)^3 =
\frac{D^3}{a_0^3} \left\{ 1 + \frac{3}{2} \left( \frac{H_0 D}{c} \right)
\right. \nonumber \\
&& \left. + \left( \frac{7}{4} + \frac{q_0}{2} \right) \left( \frac{H_0 D}{c} \right)^2
+ O \left[ \left( \frac{H_0 D}{c} \right)^3 \right] \right\}~.
\eea
Using (\ref{r0ser}), (\ref{r01}), and (\ref{r03}) we get
\bea
&& r_{0}(D) = \frac{c}{a_0 H_0} \left\{ \left( \frac{H_0 D}{c} \right)
+ \frac{1}{2} \left( \frac{H_0 D}{c} \right)^2 \right. \nonumber \\
&& \left. + \frac{1}{6}
\left( 2 + q_0 - \Omega_{K0} \right) \left( \frac{H_0 D}{c}
\right)^3 \right. \nonumber \\
&& \left. + \frac{1}{4} \left( q_0 + \frac{j_0}{6} + 1 -
\Omega_{K0} \right) \left( \frac{H_0 D}{c} \right)^4 + \right.
\nonumber \\
&& \left. \left[ \frac{1}{5} \left(1 + \frac{j_0}{3} + \frac{q_0^2}{4} +
\frac{3}{2} q_0 + \frac{k_0}{24} \right) - \frac{1}{12} \left(q_0
+ \frac{7}{2} \right) \Omega_{K0} \right] \right. \nonumber \\
&& \left. \times \left( \frac{H_0 D}{c}
\right)^5 + O\left[ \left( \frac{H_0 D}{c} \right)^6 \right]
\right\}~.
\eea
In view of the definition (\ref{lum}), it is useful to express $r_0$
as a function of redshift $z$, using formula (\ref{D(z)}), i.e.,
\bea
\label{r0z}
&& r_{0}(z) =  \frac{c}{a_0H_0} \times \nonumber \\
&&  \left\{ z - \frac{1}{2} (q_0 + 1) z^2
+ \left[ \frac{q_0^2}{2} + \frac{1}{3} (2q_0 + 1) - \frac{j_0}{6}
- \frac{\Omega_{K0}}{6} \right] z^3 \right. \nonumber \\
&& \left. + \left[ \frac{5}{12} q_0 j_0 - \frac{k_0}{24} + \frac{9}{24} j_0
- \frac{5}{8} q_0^3 - \frac{3}{4} \left( \frac{3}{2} q_0^2 + q_0 +
\frac{1}{3} \right) \right. \right. \nonumber \\
&& \left. \left. + \frac{1}{4} \left( q_0 + 1 \right) \Omega_{K0}
\right] z^4 + O(z^5) \right\}~.
\eea
Finally, from (\ref{lum}) and (\ref{r0z}), we have the
fourth-order in redshift $z$ formula for the luminosity distance
\bea
\label{lumgeneralz4}
&&D_{L}(z) = \frac{cz}{H_0} \times \nonumber \\
&& \left\{1 + \frac{1}{2} (1-q_0)z +
\frac{1}{6} \left[q_0(3q_0 + 1) - (j_0 + 1) - \Omega_{K0} \right]z^2
\right. \nonumber \\
&& \left. + \frac{1}{24} \left[ 5j_0(2q_0 + 1) - k_0 - 15q_0^2 (q_0 + 1) + 2(1 - q_0)
\right. \right. \nonumber \\
&& \left. \left. + 2 \Omega_{K0}(3q_0 + 1) \right] z^3 + O(z^4) \right\}~.
\eea
The formula (\ref{lumgeneralz4}) agrees with the third-order in $z$ formula (37) of
\cite{jerk} and with the fourth-order in $z$ formula (7) of \cite{snap}.
It also agrees with the formula (39) of \cite{AJIII} for
$\Omega_{st0} = \Omega_{g0} = \Omega_{ph0} = \Omega_{sp0} = \Omega_{bp0} = \Omega_{hp0}$.

This series expansion for the luminosity distance $D_L$ can
alternatively be obtained by using the method of Kristian and Sachs
\cite{krsachs66,ellis70,AJIII}.

Notice that one can reduce nicely the term of the third-order in
redshift $z$ by using the formulas (\ref{j0elim}) and (\ref{Ka}) to get
(we dropped the fourth-order term which can also be reduced using
these expressions together with (\ref{k0elim}))
\bea
\label{lumgeneralz3}
D_L &=& \frac{cz}{H_0} \left\{1 + \frac{1}{2} (1-q_0)z + \right.\\
&&\left.
\frac{1}{6} \left[3q_0(q_0 + 1) - 15 \Omega_{st0} - 10 \Omega_{g0} - 6 \Omega_{r0}
 \right. \right. \nonumber \\
&& \left.\left. - 3 \Omega_{m0} - \Omega_{s0} - \Omega_{ph0} - 3 \Omega_{sp0}
\right. \right. \nonumber \\
&& \left.\left.
- 6 \Omega_{bp0} - 10 \Omega_{hp0} \right]z^2 + O(z^3) \right\}~.\nonumber
\eea

\subsection{Luminosity distance - some special cases}

Let us now start with the exact luminosity distance formula in the two simple cases
of only standard matter (\ref{aforstan}) or
phantom (\ref{aforphan}), respectively, which read as (compare \cite{AJIII})
\be
D_L = \frac{2}{3\gamma -2} \frac{c}{H_0} \left[(1+z) -
(1+z)^{\frac{4-3\gamma}{2}}\right]~, \hspace{0.3cm} (\gamma \neq 2/3)~,
\ee
and
\be
D_L = \frac{2}{3\mid \gamma \mid + 2} \frac{c}{H_0} \left[(1+z) -
(1+z)^{\frac{3\mid \gamma \mid +4}{2}}\right]~.
\ee

In the special case of Section III.A, if we apply the fact that
the comoving distance $\chi$ is related to the cosmological time according to
(\ref{geodeq}), and change the variable $x = (a/a_0)^2$, we obtain
\begin{equation}
    \left(\frac{c}{2H_0a_0}\right)^2 \left(\frac{\ud x}{\ud\chi}\right)^2 = x W(x), \label{dxdchi}
\end{equation}
which can again be transformed into the Weierstrass equation, as $W$ is of degree three only.
Consequently, we can obtain the following
\bea
    &&\wp(\frac{2H_0a_0}{c}\chi) \equiv  P_{\chi} =  \\
    &&= \frac{1}{12z^2(2+z)^2} \left\{
      \K(z^4+4z^3+10z^2+12z+6) \right. \nonumber\\
       &&+ 3[\r(1+z)^2(2+2z+z^2)+\L(2+2z+z^2)+2\sp] \nonumber \\
       &&+ \left. 6\sqrt{\r(1+z)^6+\K(1+z)^4+\L(1+z)^2+\sp} \right\} \nonumber, \label{p_chi}
\eea
where we have used the definition of redshift (\ref{z}). The function $\wp$ is not the
same as in an appropriate cosmological solution. The invariants here are
\bea
    g_2 &=& \frac{1}{12}\K^2 - \frac14\L\r, \nonumber\\
    g_3 &=& \frac{1}{48}\L\K\r - \frac{1}{216}\K^3 - \frac{1}{16}\sp\r^2.
    \nonumber
\eea
Substituting (\ref{p_chi}) into (\ref{lum}), we get
\begin{equation}
    D_L = \frac{(1+z)c}{H_0}S_{\k}[\frac12\wp^{-1}(P_{\chi})], \label{DL}
\end{equation}
where the new subscript of $S$ indicates that instead of $K$ we need
to use the related density $\k$.

Since we know that $x=0$ is necessarily a root of the right hand side of equation (\ref{dxdchi}),
we can easily obtain a special case with a double root, if we put $\r=0$. The Weierstrass
function will then reduce to a trigonometric one, and the luminosity distance
will become
\begin{equation}
    D_L = \frac{c(1+z)}{H_0}S_{\k}\left[\frac{1}{\sqrt{-\K}}
        \mathrm{arccot}\left(\sqrt{-\frac{4P_{\chi}}{\K}-\frac23}\right)\right].
\end{equation}
If we assume that $\Omega_{cs0}=0$, or in other words $\K = -\k$, the above formula
is further simplified to
\begin{equation}
    D_L = \frac{c(1+z)}{H_0\sqrt{4P_{\chi}+\frac13\k}}.
\end{equation}

For practical application, it might also be useful to consider the series expansion of
the formula (\ref{DL}), which reads as
\bea
D_L &=&
\frac{cz}{H_0}\left\{ 1 + \frac{2-\K-2\r+\sp}{2}z \right. \nonumber \\
&+& \frac16\big[ 3\K^2-12\r-\k + 3(2\r-\sp)^2 \nonumber \\
&-& \left. 2\K(2-6\r+3\sp) \big]z^2 + \mathcal{O}(z^3) \right\},
\eea
and it is in agreement with our general formula
(\ref{lumgeneralz3}) after rearranging the terms by using
(\ref{Ka})-(\ref{j0elim}).

\subsection{$O(z^4)$ redshift-magnitude relation $m(z)$ formula}

As for the apparent magnitude $m$ of the source we have
\be
\label{mboldef}
m = 5\log_{10}{D_L} + M~,
\ee
where $M$ is the absolute magnitude. We will expand the apparent magnitude
into series using the luminosity distance expansion (\ref{lumgeneralz4}).
Following the Refs. \cite{krsachs66,ellis70}, in order to adopt to the standard
astronomical notation for the magnitude we first write (\ref{mboldef}) as
\be
m - M = \frac{5}{2} \log_{10}{D_{L}^{~2}}~.
\ee
Using (\ref{lumgeneralz4}), one easily gets
\bea
&& m - M = 5\log_{10}(cz) - 5\log_{10}{H_0} \nonumber \\
&& + \frac{5}{2}
\log_{10} \left\{ 1 + (1-q_0) z \right. \nonumber \\
&& \left. + \frac{1}{3} \left[\frac{15}{4}
q_0^2 - \frac{q_0}{2} - \left(j_0 + \frac{1}{4} \right) -
\Omega_{K0} \right] z^2 \right.  \\
&& \left. + \frac{1}{12} \left[ j_0 (10 q_0 + 3) -
k_0 - 3q_0^2 (5q_0 + 3) + 6 q_0 \Omega_{K0} \right] z^3 \right. \nonumber \\
&& \left. + O(z^4)
\right\}~. \nonumber
\eea
This can further be expanded using the properties of logarithms as
\be
\frac{5}{2} \log_{10} {g(z)} = \frac{5}{2}
\frac{\ln{g(z)}}{\ln{10}} = \left(\frac{5}{2} \log_{10}{e} \right)
\ln{g(z)}~,
\ee
where $g(z)$ is a given function of redshift $z$, to get
\bea
\label{m(z)3}
&& m-M = 5\log_{10}{(cz)} - 5\log_{10}{H_0}
\left(\frac{5}{2} \log_{10}{e} \right) \nonumber \\
&& \left\{ (1-q_0) z + \frac{1}{3} \left[\frac{q_0}{2}
\left(\frac{9}{2} q_0 + 5 \right) - j_0 - \frac{7}{4} -
\Omega_{K0} \right] z^2 \right. \nonumber \\
&& \left. \frac{1}{24} \left[ 2j_0 \left(8q_0 + 5 \right)
- 2k_0 - q_0 \left(7q_0^2 + 11 q_0 + 23 \right) + 25 \right. \right.
\nonumber \\
&& \left. \left. + 4 \Omega_{K0} \left(2 q_0 + 1 \right)
\right]z^3 + O(z^4) \right\}~.
\eea
From (\ref{m(z)3}) it is clear that the jerk appears in the second
order of the expansion and the "kerk" appears in the third order
of this expansion. Also, if $O(z^3)$ term is dropped, this formula agrees with the formula
(41) of Ref. \cite{AJIII}, i.e.,
\bea
m &=& M - 5 \log_{10}{H_0} + 5 \log_{10}{(cz)} \\
&+& (2.5 \log_{10}{e}) \left\{(1-q_0) z +
\left[\frac{1}{4} (3q_0 + 1)(q_0 - 1) - \right. \right. \nonumber \\
&&\left. \left.
\frac{2\sigma_{s0}}{3} -
\frac{2\Lambda_{0}}{3H_0^2} \right] z^2 + O(z^3) \right\}, \nonumber
\eea
in which all but radiation, dust, $\Lambda$-term, and cosmic
string terms are neglected. This can easily be checked after the
application of the relations (\ref{Ka}, (\ref{Lambda0}), and (\ref{j0elim}).
The most general, second order $O(z^2)$ relation, for all the
types of matter admitted in this paper reads as
\bea
&&m = M - 5 \log_{10}{H_0} + 5 \log_{10}{(cz)} \\
&+& (2.5 \log_{10}{e}) \left\{(1-q_0) z +
\left[\frac{3}{2} \left( \frac{q_0}{2} + 1 \right) - \frac{1}{4} \right. \right. \nonumber \\
&&\left. \left. - 5 \Omega_{st0} - \frac{10}{3} \Omega_{g0} - 2
\Omega_{r0} - \Omega_{m0} - \frac{1}{3} \Omega_{s0} \right. \right. \nonumber \\
&& \left. \left. - \frac{1}{3}
\Omega_{ph0} - \Omega_{sp0} - 2 \Omega_{bp0} - \frac{10}{3}
\Omega_{hp0} \right] z^2 + O(z^3) \right\}~.\nonumber
\eea

\subsection{Angular Diameter}

This quantity is given by:
\begin{equation}
    \theta = \frac{d(1+z)^2}{D_L},
\end{equation}
where $d$ is the linear size of a given object. Using the formula
(\ref{geodeq}) we can consider the minimum of $\theta(z)$, which is a solution
of the equation
\begin{equation}
    \frac{S_K(\chi)}{S_K'(\chi)} = (1+z)\frac{\ud\chi}{\ud z}.
\end{equation}
In general this is a transcendental relation, which is impossible to solve in a suitably
closed form. However, in a special case of $\r=\Omega_{sc0}=0$, this equation
admits an algebraic solution, as it simplifies to:
\begin{equation}
    12P_{\chi}-2\k = 3(1+z)^2W\left[\frac{1}{(1+z)^2}\right].
\end{equation}

Finally, the angular size of a galaxy in a one-component
standard (\ref{aforstan}), and phantom (\ref{aforphan}) matter models are
($d$-linear size of this galaxy)
\be
\Delta \theta_1 = \frac{H_0 (3\gamma-2)}{2c} \frac{d(1+z)^2}{(1+z)
- (1+z)^{\frac{4-3\gamma}{2}}}~,\hspace{0.3cm} (\gamma \neq 2/3)~,
\ee
and
\be
\Delta \theta_1 = \frac{H_0 (3\mid \gamma \mid + 2)}{2c} \frac{d(1+z)^2}{(1+z)
- (1+z)^{\frac{4+3\mid \gamma \mid}{2}}}~,\hspace{0.3cm} (\gamma \neq
2/3)~.
\ee
In the former case, there is a minimum at \cite{AJIII}
\bea
z_{min} &=& \left(\frac{2}{3\gamma}\right)^{\frac{2}{2-3\gamma}} -
1~, \hspace{0.3cm} (\gamma \neq 2/3); \\
z_{min} &=& e-1~, \hspace{0.3cm} (\gamma = 2/3)~;\hspace{0.3cm}
z_{min} \to \infty \hspace{0.3cm} (\gamma =0)~,\nonumber
\eea
while for the latter (phantom) the minimum does not appear at all.

\subsection{Source Counts}

We can also calculate the number of sources with redshifts from the
interval $z$, $z+dz$ and the density of sources $n(z)$
\begin{equation}
    N = 4\pi n(z) a_0^3 S_K^2[\chi(z)]\frac{\ud\chi}{\ud z},
\end{equation}
can be rewritten using (\ref{dxdchi}) as
\begin{equation}
    N = 4\pi n(z) \left(\frac{c}{H_0}\right)^3 \frac{S_{\k}^2[\frac12\wp^{-1}(P_{\chi})]}
        {(1+z)^2\sqrt{W[(1+z)^{-2}]}},
\end{equation}
or, assuming constant $n(z)=n=$ const., it can be expanded into
\begin{equation}
    N = 4\pi n \left(\frac{c}{H_0}\right)^3 \bigg\{ z^2 - 2(\K+2\r-\sp)z^3 + \mathcal{O}(z^4) \bigg\}.
\end{equation}
The $O(z^4)$ formula for radiation, matter, strings and
$\Lambda$-term models was given in \cite{AJIII} as
\bea
\label{Nz}
N(z) &=& 4\pi n \left(\frac{2\Omega_{r0} + \frac{3}{2} \Omega_{m0} + \Omega_{s0} - q_0 -
1}{k}\right)^{\frac{3}{2}}  \\
&\times& \left\{z^2 - 2 (q_0 + 1)z^3 + \frac{1}{12}
\left[(q_0+1)(37q_0+31) \right. \right. \nonumber \\
&-&  \left. \left. 48\Omega_{r0}
- 21\Omega_{m0} - 4 \Omega_{s0} \right]z^4
+ O(z^5) \right\} .\nonumber
\eea

However, both angular diameter and source counts tests, in view of supernovae data
\cite{supernovae}, are not
so precise as redshift-magnitude relation test, which is the reason we do not
investigate them in so much detailed way, as we do for the redshift-magnitude.

\section{Conclusion}
\setcounter{equation}{0}

We have studied more general phantom ($p < - \varrho$) cosmological models
(cf. \cite{FRWelliptic,phantom1}), which include all the types of both
phantom and standard matter from the barotropic index range $-7/3 < w <
1$.

We have found that there are various interesting possibilities of
the evolution, depending on the matter content in the universe.
Since phantom cosmologies allow both standard Big-Bang ($a \to 0$, if $t \to
0$) and phantom-driven Big Rip ($a \to \infty$, if $t \to t_0$)
singularities, then the set of possible types of evolution
enlarges. From the observational point of view the most interesting
is the Big-Bang to Big-Rip (BB-BR) type of evolution which is strongly
supported by the current supernovae data \cite{supernovae}.
However, a lot of other interesting theoretical options related to this are also
possible. For example, there exists an unstable static universe
(US) which, if perturbed, may go into one of the two monotonic
universes - one of them is monotonic towards a Big-Bang (MBB type) and another
one is monotonic towards a Big-Rip (MBR type). More hybrid solution is that
there exist two unstable static universes, if perturbed, except
for MBB and MBR options, there is a possibility to have a solution
which is monotonic from one US solution to the other (MUS-MUS type).
In fact, one or the two monotonic solutions may have a bounce - it is of
monotonic non-singular
(MNS) type. Stable static solutions (SS) are also possible.
The standard Big-Bang to Big-Bang (BB-BB) type of solutions in
phantom cosmologies may also be replaced by Big-Rip to Big-Rip
(BR-BR) type. Finally, there exist a large class of non-singular
oscillating (ONS) type of solutions - in some cases oscillations
are allowed for the two different ranges of the values of the
scale factor.

We have also studied observational characteristics of phantom
cosmologies. We enlarged the set of the dynamical parameters of
cosmological models to include (except for the Hubble constant and
the deceleration parameter) the jerk and the "kerk". The
"kerk" is defined by the fourth order derivative of the scale factor.
This allowed us to write down the luminosity distance
relation $D_{L}(z)$ and the redshift-magnitude relation $m(z)$ up to the fourth
order term in the series expansion of redshift $z$. We conclude that jerk appears
in the third order
of the expansion while the "kerk" appears in the
fourth order of the expansion, as expected. Both jerk and "kerk" can be useful as
state-finders - the parameters which can be helpful in
determination of the equation of state of the cosmic fluid \cite{jerk,snap}.
Of course these considerations for state-finders are valid, provided we assume a
barotropic (or analytic) type of the equation of state and do
not apply in the case of sudden future singularities which do not
tight $p$ and $\varrho$ \cite{barrow04,inhosfs}.

Finally, we have discussed other observational tests for
phantom models such as angular diameter test and source counts. It
is interesting that for one fluid phantom models the angular
diameter minimum does not appear at all - this is in clear
contrast to a standard matter models where the minimum can be
an important characteristic of these models.

As a further step towards the more negative values of pressure in
cosmology (despite some objections \cite{kaloper}), phantom cosmologies
are a viable completion of standard cosmologies which may solve cosmological
puzzles.

\section{Acknowledgements}

M.P.D. acknowledges the support from the Polish Ministry of Science and Computing
grant No 1P03B 043 29 (years 2005-2007).

\appendix

\section{General classification of the models which lead to
a cubic polynomial in the canonical equation}

We take into account the canonical equation
\begin{equation}
        \left(\frac{\ud x}{\ud\tau}\right)^2 = a_3x^3+a_2x^2+a_1x+a_0 = W(x) \label{basic}
\end{equation}
together with the constraint
\begin{equation}
        a_3+a_2+a_1+a_0=1, \label{cond}
\end{equation}
for all non-static solutions. The reason for this is the normalization of
an independent variable, so that for a particular value of $x$ its derivative is equal
to a fixed constant. Here we selected $W(1)=1$. This allows easier classification,
as it directly corresponds to a shape of the polynomial $W(x)$. Since the parts of
the curve $\{(W(x),x):\;W(x)\geq 0\}$ will in general be disconnected,
a physically significant solution will be the one for which $x$ passes through $1$.

In the case of a static solution, $W(x)=0$ at all times, and such a normalization is
impossible.

The behaviour of the solutions depends primarily on the roots of the equation $W(x)=0$.
As we are only interested in real and positive solutions, the position of the roots,
determine the intervals where $W(x)$ is positive, i.e., where physically significant
evolution can take place. To this end, we introduce the following notation:
\bea
        e_i&:& \quad W(x)=a_3(x-e_1)(x-e_2)(x-e_3),\nonumber\\
        \Delta &:=& (e_1-e_2)^2(e_2-e_3)^2(e_3-e_1)^2 \nonumber\\
        &\phantom{:}=& \frac{a_2^2a_1^2}{a_3^4}-4\frac{a_1^3}{a_3^3}-4\frac{a_2^3 a_0}{a_3^4}
                +18\frac{a_2 a_1 a_0}{a_3^3}-27\frac{a_0^2}{a_3^2}.\nonumber \\
        g_2&:=& \frac{1}{12}a_2^2-\frac14 a_3a_1, \nonumber\\
        g_3&:=& \frac{1}{48}a_3a_2a_1-\frac{1}{216}a_2^3-\frac{1}{16}a_3^2a_0,
        \nonumber
\eea
where $e_i$ are the roots of $W(x)$, $\Delta$ is the discriminant,
and $g_2$, $g_3$ are invariants \cite{abramovitz}.
For static solutions, the appropriate quantities are those of equation
(\ref{basic}) before it is normalized.

The invariants $g_2$ and $g_3$ appear when equation (\ref{basic}) is transformed
by
\be
x=\frac{4}{a_3}v-\frac{a_2}{3a_3}~.
\ee
It is then simplified to
\begin{equation}
        \left(\frac{\ud v}{\ud\tau}\right)^2 = 4v^3-g_2v-g_3 = V(v), \label{basic_w}
\end{equation}
and can immediately be solved by means of the Weierstrass elliptic $\wp$
function, i.e.,
\begin{equation}
        v = \wp(\tau-\tau_0;g_2,g_3).
\end{equation}

\noindent Now, we consider the following cases:

\subsection{$a_3>0$}

\subsubsection{$\Delta>0$}

Necessarily, there exist three distinctive, real roots of the polynomial $W(x)$ in this case.
Arranging them so that $e_1>e_2>e_3$, the condition (\ref{cond}) implies $e_3<1<e_2$ or
$e_1<1$. Accordingly, there are two regions of admissible values of $x$:
$e_3\leq x_-\leq e_2$ and $e_1\leq x_+$, respectively. An appropriate general solution is
\begin{equation}
    x_{\phantom{-}}
        = \frac{\wp'(\tau-\tau_0)+ a_{321}
        [\wp(\tau-\tau_0)-\frac{1}{12}(3a_3+a_2)]+\frac14 a_3}
        {2[\wp(\tau-\tau_0)-\frac{1}{12}(3a_3+a_2)]^2},
\end{equation}
with
\bea
a_{321} = \frac12(3a_3+2a_2+a_1)~,\nonumber
\eea
or, for particular cases
\bea
        x_- &= \frac{4}{a_3}\wp(\tau-\tau_0+\omega_3)-\frac{a_2}{3a_3} \nonumber\\
            &= e_3+\frac{3a_3(e_3-e_1)(e_3-e_2)}{12\wp(\tau-\tau_0)-(2e_3-e_1-e_2)} \nonumber\\
            &= e_3+\frac{6W'(e_3)}{24\wp(\tau-\tau_0)-W''(e_3)}, \label{xm1a}\\
        x_+ &=
        \frac{4}{a_3}\wp(\tau-\tau_0)-\frac{a_2}{3a_3}.\label{xp1a}
\eea
Here $\tau_0$ is a real constant, and $\omega_3$ is a purely imaginary half-period of
$\wp$, given by
\bea
        \omega_3=\int_{-\infty}^{e_3}\frac{\ud t}{\sqrt{4t^3-g_2t-g_3}}.\nonumber
\eea
The solution $x_-$ is a non-singular solution,
if $e_3 > 0$, while $x_+$ is singular, when $\tau-\tau_0 = 2 n\omega_1,\; n\in {\bf Z}$
(Fig.\ref{fign311}).

Both of the solutions (\ref{xm1a}) and (\ref{xp1a}) are periodic with the real period
\bea
\omega_1=\int_{e_1}^{\infty}\frac{\ud t}{\sqrt{4t^3-g_2t-g_3}}.\nonumber
\eea

\begin{figure}[h]
\includegraphics[angle=0,scale=.46]{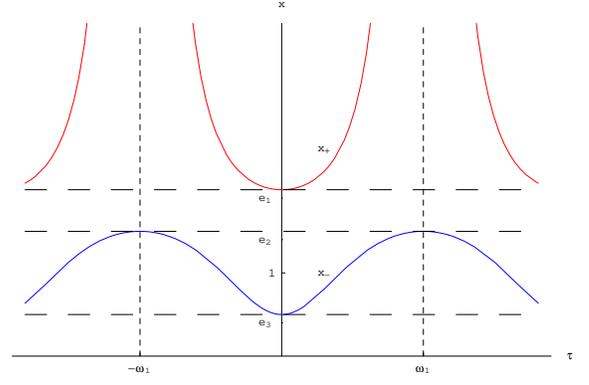}
\caption{Case A.1.a -- three different roots.
There is an oscillating non-singular (ONS) solution $x_-$ given by (\ref{xm1a}) and a singular
solution (of Big-Rip to Big-Rip (BR-BR) type) $x_+$ given by (\ref{xp1a}).}
\label{fign311}
\end{figure}

\subsubsection{$\Delta=0$}

Here, there is a double or a triple root. In order to distinguish between these two possibilities,
we use the equation (\ref{basic_w}). The roots of the polynomial $V(v)$:
$\widetilde{e}_1$, $\widetilde{e}_2$, $\widetilde{e}_3$, must satisfy:
$\widetilde{e}_1+\widetilde{e}_2+\widetilde{e}_3=0$. Therefore a triple root of $W(x)=0$
corresponds to a triple root $v=0$ of $V(v)$, with $g_2=g_3=0$.

Let us investigate the former case first.
Obviously, there must exist another real root, but not necessarily positive.
That is, $W(x)=a_3(x-e_1)(x-e_2)^2$.

If the double root is a smaller one, then because $a_3>0$, we obtain a stable, static
solution $x=e_1$. If the opposite is true, the same static solution is unstable.

For $e_2<e_1$ ($e_1 < 1$) we have the following solution with $x\geq e_1$
(Fig. \ref{fign312})
\begin{equation}
\label{x01a}
        x_0 = e_1+(e_2-e_1)\tan^2\left[\frac{\sqrt{a_3(e_1-e_2)}}{2}(\tau-\tau_0)\right],
\end{equation}
which is periodic with a half-period
\bea
\omega_1=\frac{\pi}{\sqrt{a_3(e_1-e_2)}}~.\nonumber
\eea

\begin{figure}[h]
\includegraphics[angle=0,scale=.46]{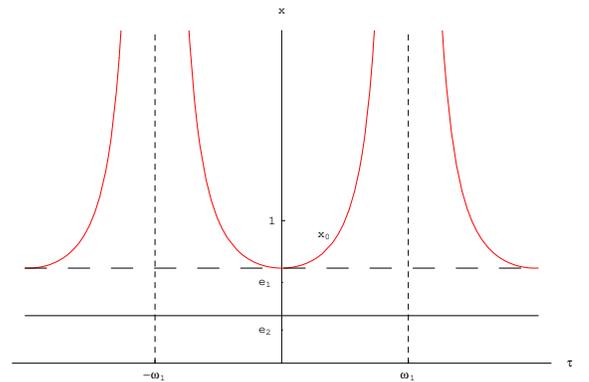}
\caption{Case A.1.b - the smaller root $e_2$ is double.
There is a stable static solution (SS) $x=e_1$ and a solution $x_0$ (BR-BR type) given by
(\ref{x01a}).}
\label{fign312}
\end{figure}

In fact, this is valid for both $e_2<e_1$ and $e_2>e_1$. However, when $e_2>e_1$, the solution
is no longer periodic, and, upon eliminating the imaginary part of $\tau_0$, it is more
convenient to write it as (Fig. \ref{n312b})
\bea
        x_- & =&
        e_1+(e_2-e_1)\tanh^2\left[\frac{\sqrt{a_3(e_2-e_1)}}{2}(\tau-\tau_0)\right],\nonumber
        \\ &&\text{when $e_1 < 1 < e_2$}\label{xm1b}\\
        x_+ & =&
        e_1+(e_2-e_1)\coth^2\left[\frac{\sqrt{a_3(e_2-e_1)}}{2}(\tau-\tau_0)\right],\nonumber
        \\ &&\text{when $e_2 < 1$.\label{xp1b}}
\eea

\begin{figure}[h]
\includegraphics[angle=0,scale=.46]{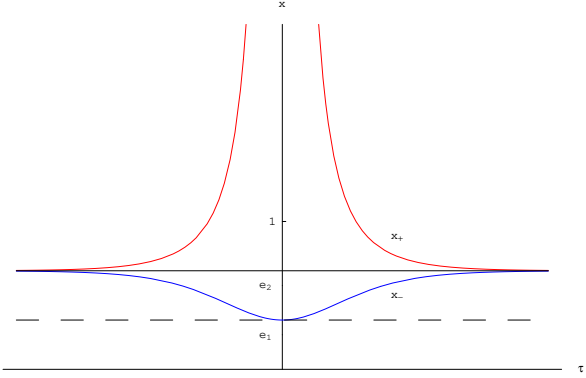}
\caption{Case A.1.b -- the greater root $e_2$ is double. There is an
unstable static solution (US) $x=e_2$ and the two monotonic solutions $x_{\pm}$ given by
(\ref{xm1b}) and (\ref{xp1b}). One of them is monotonic non-singular (MNS)
and another monotonic to a Big-Rip singularity (MBR).}
\label{n312b}
\end{figure}

In terms of the coefficients of the polynomial $W(x)$, it is possible to discern between
these two possibilities, as clearly seen from the sign of $W(x_1)$. The value at the point
of inflection is positive when $e_2 > e_1$, and negative otherwise. This relation
can be precisely written as: $16g_3=-a_3^2W(x_1)=\frac{2}{27}a_3^3(e_1-e_2)^3$.

If there is only one triple root $e_2$, there is an unstable static solution
$x=e_2$, and an asymptotic solution (Fig. \ref{x01b})
\begin{equation}
\label{solx01b}
        x_0 = e_2+\frac{4}{a_3(\tau-\tau_0)^2}.
\end{equation}
That is, a ``small perturbation'' causes the monotonic evolution. Of course, if $e_2<0$,
only the latter solution remains valid and it is of BB-BR type.

\begin{figure}[h]
\includegraphics[angle=0,scale=.46]{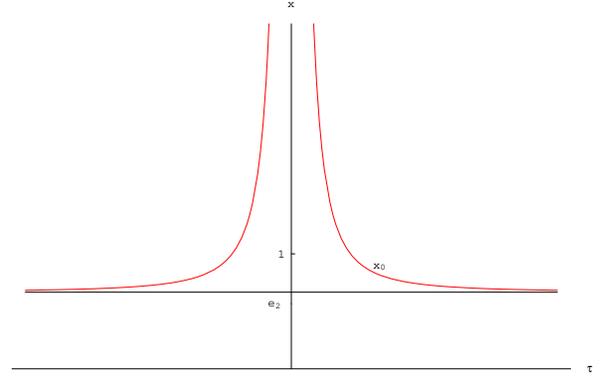}
\caption{Case A.1.b -- one triple root. There is an unstable
static solution (US) $x=e_2>0$ and a monotonic solution (MBR) $x_0$ given by (\ref{solx01b}).}
\label{x01b}
\end{figure}

\subsubsection{$\Delta<0$}

Only one real root exists in this case, and because of (\ref{cond}), it must be
smaller than unity. Then, there is only one type of solution
\begin{equation}
\label{x01c}
        x_0 = \frac{4}{a_3}\wp(\tau-\tau_0)-\frac{a_2}{3a_3},
\end{equation}
which is singular when the argument
$\tau-\tau_0 = 4 n {\mathcal Re}(\omega_1),\; n\in {\bf Z}$ (Fig. \ref{n313}).

\begin{figure}[h]
\includegraphics[angle=0,scale=.46]{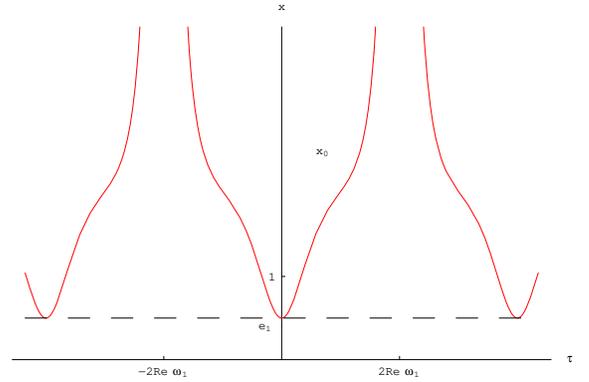}
\caption{Case A.1.c -- one real root exists only.
The solution $x_0$ is given by (\ref{x01c}). It is a BR-BR type of solution with
no Big-Bang (BB) singularity.}
\label{n313}
\end{figure}

\subsection{$a_3<0$}

\subsubsection{$\Delta>0$}

This case is, in fact, the same as $a_3 >0$, with an interchange of possible
regions of the evolution. This gives rise to a periodic solution oscillating
between the two greater roots $e_2 \leq x \leq e_1$ (Fig. \ref{n321})
\begin{equation}
\label{xp2a}
            x_+ = e_1+\frac{3a_3(e_1-e_2)(e_1-e_3)}{12\wp(\tau-\tau_0)-(2e_1-e_2-e_3)},
\end{equation}
where the roots have been arranged as before.

Another solution, which in the previous case A.1 was singular, now is also bounded
$0 \leq x \leq e_3$, and is given by (Fig. \ref{n321})
\begin{equation}
\label{xm2a}
            x_- = \frac{4}{a_3}\wp(\tau-\tau_0)-\frac{a_2}{3a_3}.
\end{equation}

\begin{figure}[h]
\includegraphics[angle=0,scale=.46]{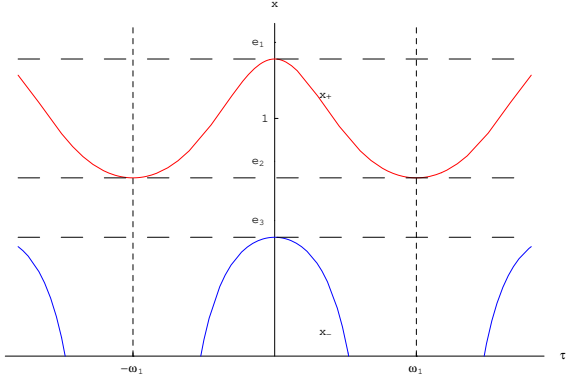}
\caption{Case A.2.a -- three different roots. There is an oscillating non-singular
solution (ONS) $x_+$ given by (\ref{xp2a}), and a singular solution (of Big-Bang to
Big-Crunch (BB-BC) type) $x_-$ given by
(\ref{xm2a}).}
\label{n321}
\end{figure}

\subsubsection{$\Delta=0$}

As in the case A.1.b of the $a_3>0$ case, we can either have a double or a triple root.
Also, we can distinguish between the two double root subcases, using the same formulas as
before.

If the root is only double we have a situation similar to what we had before.
It can be a greater root,
in which case the other root must satisfy $e_1>1$, for only then we could have $W(1)=1$. We
can only have a stable, static solution $x=e_2$, and another one for
$0\leq x \leq e_1$, given by (Fig. \ref{n322b}):
\begin{equation}
\label{x02b}
        x_0 = e_1 + (e_1-e_2)\tan^2\left[\frac{\sqrt{|a_3|(e_2-e_1)}}{2}(\tau-\tau_0)\right].
\end{equation}
(Again, this is valid for all double-root solutions, but $\tau_0$ might be complex.)
\begin{figure}[h]
\includegraphics[angle=0,scale=.46]{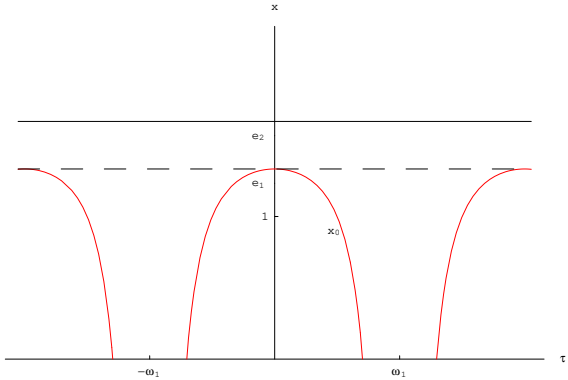}
\caption{Case A.2.b -- the greater root $e_2$ is double.
There is a stable static solution (SS) $x=e_1$ and a BB-BC type singular solution $x_0$ given by
(\ref{x02b}).}
\label{n322b}
\end{figure}

On the other hand, when $e_2<e_1$, which requires $e_1>1$, there are three possible
solutions. One is an unstable static (US) solution $x=e_2$. The other two describe the motion
between $0\leq x \leq e_2$ (asymptotic to the previous one) and $e_2\leq x\leq e_1$.
They are given by (Fig. \ref{n322a})
\bea
        x_- & =& e_1 + (e_2-e_1)\coth^2\left[\frac{\sqrt{|a_3|(e_1-e_2)}}{2}(\tau-\tau_0)
        \right], \nonumber \\
                &&\text{when $1 < e_2$}\label{xm2b}\\
        x_+ & =& e_1 + (e_2-e_1)\tanh^2\left[\frac{\sqrt{|a_3|(e_1-e_2)}}{2}(\tau-\tau_0)
        \right], \nonumber \\
                &&\text{when $e_2 < 1 < e_1$.\label{xp2b}}
\eea

\begin{figure}[h]
\includegraphics[angle=0,scale=.46]{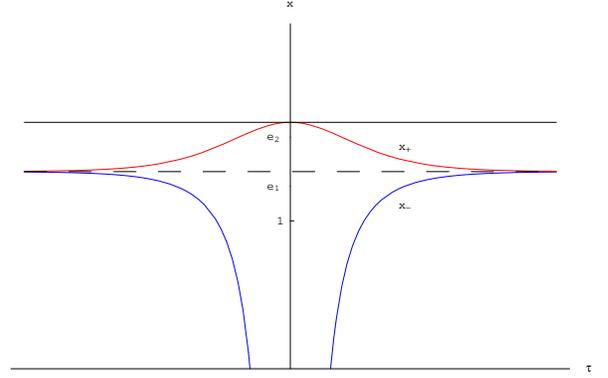}
\caption{Case A.2.b -- the smaller root $e_2$ is double.
There is an unstable static (US) solution $x=e_1$ and the two monotonic solutions $x_{\pm}$
given by (\ref{xm2b}) and (\ref{xp2b}). The former is of monotonic non-singular (MNS)
type and the latter is monotonic Big-Bang type (MBB).}
\label{n322a}
\end{figure}

When $e_2$ is a triple root, there is an unstable, static solution $x=e_2$, and a monotonic
solution in the form (Fig. \ref{n322c})
\begin{equation}
\label{x02btriple}
        x_0 = e_2+\frac{4}{a_3(\tau-\tau_0)^2}.
\end{equation}

In order for condition (\ref{cond}) to be satisfied, we must have $e_2>1$, as $W(x)$
is positive only for $x<e_2$.

\begin{figure}[h]
\includegraphics[angle=0,scale=.46]{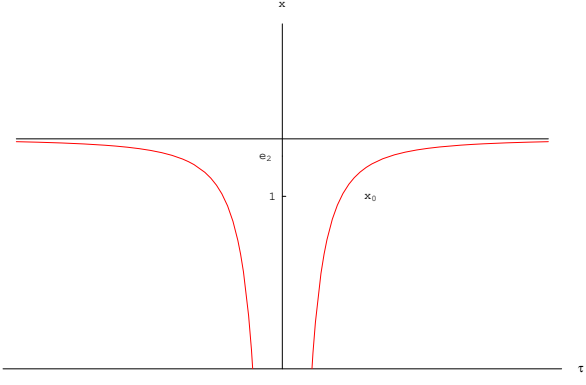}
\caption{Case A.2.b -- one triple root.
There is an unstable static (US) solution $x=e_2$ and a monotonic Big-Bang (MBB)
solution $x_0$ given by (\ref{x02btriple}).}
\label{n322c}
\end{figure}

\subsubsection{$\Delta<0$}

As in the case 1.c, there is only one real root $e_1$, only this time it limits the
values of $x$ to $0 \leq x \leq e_1$.
The solution is singular of Big-Bang to Big-Crunch (BB-BC) type (Fig. \ref{n323})
\begin{equation}
\label{x02c}
        x_0 = \frac{4}{a_3}\wp(\tau-\tau_0)-\frac{a_2}{3a_3},
\end{equation}

\subsection{$a_3 = 0$}

This case was studied in the Appendix A of Ref. \cite{phantom1}.

\begin{figure}[h]
\includegraphics[angle=0,scale=.46]{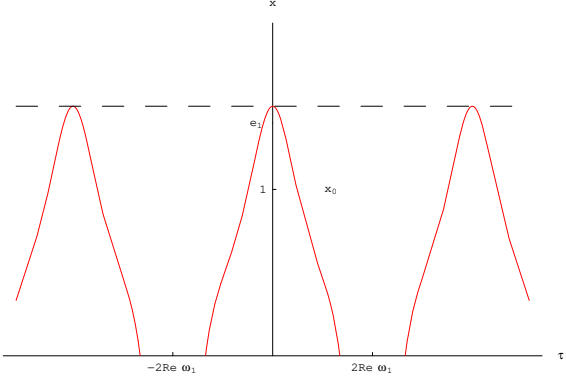}
\caption{Case 2.c -- one real root only. The solution $x_0$ is given
by (\ref{x02c}) and is of BB-BC type.}
\label{n323}
\end{figure}

\section{General classification of the models which lead to
a quartic polynomial in the canonical equation}

In this Appendix we consider the canonical equation
\begin{equation}
        \left(\frac{\ud x}{\ud\tau}\right)^2 =
        a_4x^4+a_3x^3+a_2x^2+a_1x+a_0 = W(x), \label{basic4}
\end{equation}
and impose the constraint
\begin{equation}
\label{W1}
    W(1)=a_4+a_3+a_2+a_1+a_0=1 \label{cond4}
\end{equation}
which holds with the same restrictions as in the previous case of Appendix A. In order
to make the analysis of the roots easier, we transform (\ref{W1})
to the form
\begin{equation}
    \frac{1}{a_4}W(x-\frac{a_3}{4a_4})=x^4+px^2+qx+r. \label{reduced4}
\end{equation}

Because of that, it is easier to define a set of auxiliary quantities
\bea
    \delta_{ij}=(e_i-e_j)^2~, \nonumber
\eea
where $e_i$ are the four roots of the polynomial $W(x)$, and $i,j = 1,2,3,4$. It is clear, that the original
equation (\ref{basic4}), and the transformed equation (\ref{reduced4}) have exactly the
same values of
$\delta$, and general properties of the roots remain the same. Now let $\mathcal{J}$ be
the set of all possible values of the symmetric multi-index $ij$. We can further define
the quantities:
\bea
    \sigma_6& =&\prod_{\kappa\in\mathcal{J}}\delta_{\kappa},\nonumber \\
    \sigma_5& =& \sum_{\kappa\in\mathcal{J}}
        \delta_{\kappa}^{-1}\prod_{\alpha\in\mathcal{J}}\delta_{\alpha},\nonumber\\
    \sigma_4& =&\sum_{\kappa_1,\kappa_2\in\mathcal{J}, \kappa_1<\kappa_2}
        \delta_{\kappa_1}^{-1}\delta_{\kappa_2}^{-1}\prod_{\alpha\in\mathcal{J}}
        \delta_{\alpha},\nonumber \\
    \sigma_3& =&\sum_{\kappa_1,\kappa_2,\kappa_3 \in \mathcal{J},
        \kappa_1<\kappa_2<\kappa_3} \delta_{\kappa_1}\delta_{\kappa_2}\delta_{\kappa_3},
        \nonumber\\
    \sigma_2& =&\sum_{\kappa_1,\kappa_2\in\mathcal{J},
        \kappa_1<\kappa_2}
        \delta_{\kappa_1}\delta_{\kappa_2},\nonumber \\
    \sigma_1& =&\sum_{\kappa\in\mathcal{J}} \delta_{\kappa}.
\eea
As these quantities are symmetric in the roots, they can be re-expressed by using
the coefficients $a_i$ or, equivalently, $p$, $q$ and $r$ as
\bea
    \sigma_6& =& -4p^3q^2 - 27q^4 + 16p^4r + 144pq^2r - 128p^2r^2 + 256r^3,\nonumber\\
    \sigma_5& =& -4p^5 - 18p^2q^2 - 32p^3r - 216q^2r + 192pr^2,\nonumber\\
    \sigma_4& =& 17p^4 + 48pq^2 + 24p^2r - 112r^2,\nonumber\\
    \sigma_3& =& -28p^3 - 26q^2 - 16pr,\\
    \sigma_2& =& 22p^2 + 8r,\nonumber\\
    \sigma_1& =& -8p.\nonumber
\eea
Note that $\sigma_6$ can immediately be identified as the original equation's discriminant.
Together with other $\sigma$'s, it determines the behaviour of the roots in the
following way:

\begin{enumerate}
\item{$\sigma_6 \ne 0$}\\
    There are four distinct roots, with the following subcases:\\
    \hspace*{1em}$\sigma_6<0$ -- two complex, conjugate roots, and two real roots,\\
    \hspace*{1em}$\sigma_6>0$ -- four real roots, or four complex roots, cojugate in pairs.
\item{$\sigma_6=0$, $\sigma_5 \ne 0$}\\
    There is one double, real root, and:\\
    \hspace*{1em}$\sigma_5>0$ -- two real roots,\\
    \hspace*{1em}$\sigma_5<0$ -- two complex conjugate roots.
\item{$\sigma_6=0=\sigma_5=0$, $\sigma_4 \ne 0$}\\
    There are two double roots, which for:\\
    \hspace*{1em}$\sigma_1>0$ are real,\\
    \hspace*{1em}$\sigma_1<0$ are complex, conjugate in pairs.
\item{$\sigma_6=\sigma_5=\sigma_4=0$, $\sigma_3 \ne 0$}\\
    There is a triple real root. The fourth root is also real.
\item{$\sigma_6=\sigma_5=\sigma_4=0$, $\sigma_3=0\Leftrightarrow\sigma_2=0\Leftrightarrow\sigma_1=0$}\\
    There is a quadruple root -- it is necessarily real.
\end{enumerate}

In order to distinguish between the real and complex roots when $\sigma_6>0$, we choose to
employ the resolvent polynomial of $W(x)$, which appears in the process of factorizing
a quartic polynomial into two quadratic ones
\bea
    x^4+px^2+qx+r = (x^2+kx+m)(x^2-kx+n).\nonumber
\eea
It is essentially the equation determining the value of $k$, as for $w=k^2$, the resolvent
equation is
\bea
    w^3+2pw^2+(p^2-4r)w-q^2=0.\nonumber
\eea
Now, in the case of the four real roots, such decomposition is possible in three different ways, as there
are three ways of pairing the roots. Hence, there must be three, distinct, positive solutions for $k^2$.
If, however, the roots are all complex, in order for the coefficients $k,m,n$ to be real, each root
must be paired with its conjugate. Thus, there is only one positive solution for $k^2$,
and the two negative ones. It is easy to distinguish between these possibilities, looking at the positions of the extrema
of the resolvent polynomial. In the first case, there must be the two positive ones, and in
the second, if any exist, at least one must be negative. Thus, the problem may be reduced
to an investigation of the sign of the smallest extremum:
\begin{equation}
    w_- = \frac13(-2p-\sqrt{p^2+12r}).
\end{equation}
If $p^2+12r$ is negative, and there are no extrema, only one decomposition is possible.
Note that this quantity must be non-negative in all the multiple root cases,
as degeneracy of the roots of $W$ implies multiple roots of the resolvent.

In general, when there are no multiple roots, the main equation (\ref{basic4}),
can be solved by using the substitution:
\begin{equation}
\label{Wbis}
    \frac{1}{x-e_i} = \frac{4v}{W'(e_i)} - \frac{W''(e_i)}{6W'(e_i)},
\end{equation}
which transforms it into the Weierstrass equation
\be
\dot{v}^2=4v^3-g_2v-g_3~,
\ee
with the following invariants:
\bea
    g_2& =& a_4a_0-\frac14a_3a_1+\frac{1}{12}a_2^2\nonumber\\
       & =& a_4^2(\frac{1}{12}p^2+r),\nonumber\\
    g_3& =& \frac16 a_4a_2a_0-\frac{1}{16}a_4a_1^2+\frac{1}{48}a_3a_2a_1
        -\frac{1}{16}a_3^2a_0-\frac{1}{216}a_2^3,\nonumber\\
       & =& a_4^3(-\frac{1}{216}p^3-\frac{1}{16}q^2+\frac16pr). \label{invariants4}
\eea
This, however, requires the knowledge of the roots, and usually yields too cumbersome
formulas, sometimes with explicitly imaginary coefficients.
The following solutions were possible to be simplified as a result of the
division into the special subcases. Still, the main function $\wp(\tau)$, and its half-periods,
are constructed using the invariants (\ref{invariants4}).

\subsection{$a_4>0$}

\subsubsection{$\sigma_6 \ne 0$ -- simple roots}

Choosing an integration constant by imposing that $x(0)=1$, we
obtain a general solution of the form
\bea
\label{generic4}
&&x_0 = 1+ \\
&&\frac{\wp'(\tau-\tau_0)+\frac{1}{4}(4a_4+a_3)}
{2[\wp(\tau-\tau_0)-\frac{1}{12}(6a_4+3a_3+a_2)]^2-\frac12a_4}+ \nonumber \\
&&\frac{\frac12(4a_4+3a_3+2a_2+a1)[\wp(\tau-\tau_0)-\frac{1}{12}(6a_4+3a_3+a_2)]}
{2[\wp(\tau-\tau_0)-\frac{1}{12}(6a_4+3a_3+a_2)]^2-\frac12a_4},\nonumber
\eea
which is valid for all subcases. However, the behaviour of this solution varies
greatly depending on the properties of the roots. The details are given in the
following subsections, and are clearly drawn in the figures.

\begin{center}
{\small{\it a.1. Four complex roots}}
\end{center}

The polynomial $W$ is always positive here, and from a simple, geometric
investigation it is clear that the condition (\ref{cond4}) can be satisfied. Only one solution
is present, with a possible time reversal $t\rightarrow -t$ applicable in all cases.
It reaches both Big-Bang and Big-Rip in finite times (Fig. \ref{n4111}).

\begin{figure}[h]
\includegraphics[angle=0,scale=.46]{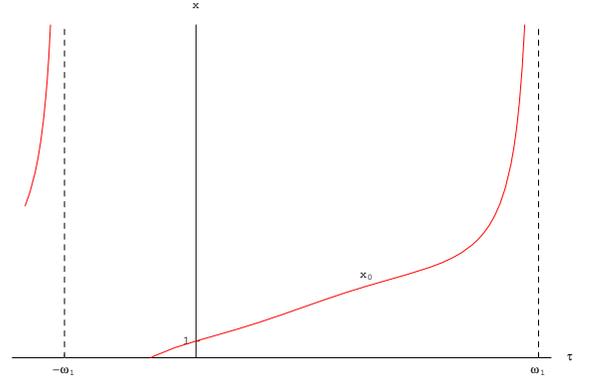}
\caption{Case B.1.a.1 -- four complex roots. The solution $x_0$ is
given by the formula (\ref{generic4}) -- the evolution starts from Big-Bang and
terminates at Big-Rip (BR-BR type).}
\label{n4111}
\end{figure}

\begin{center}
{\small{\it a.2. Two complex and two real roots}}
\end{center}

With real roots, the possible domain of $x$ is separated into two regions:
$x_-\leq e_1<e_2\leq x_+$. However, both solutions are given by the same
generic formula (\ref{generic4}. If we are interested only in physical solutions, then only
one ``branch'' is valid in each case, depending on the position of the roots.
Namely, if $1 < e_1$, we take $x_-$, and if $e_2 < 1$, we take $x_+$ (Fig. \ref{n4112}).
The solution $x_-$ starts at Big-Bang reaches the maximum and
terminates at Big-Crunch. The solution $x_+$ starts at Big-Rip
reaches the minimum and terminates at Big-Rip again.

\begin{figure}[h]
\includegraphics[angle=0,scale=.46]{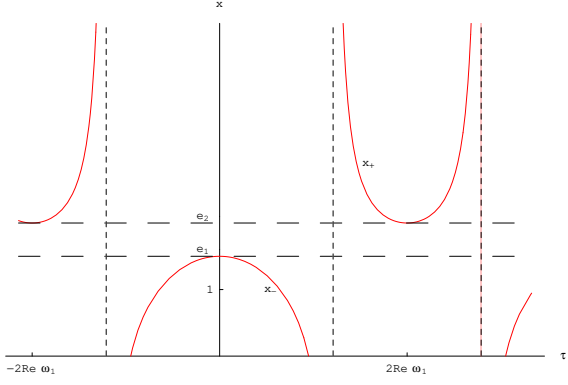}
\caption{Case B.1.a.2 -- two complex and two real roots. There are two solutions
$x_{\pm}$ both given by general formula (\ref{generic4}). The solution $x_-$
is of BB-BC type while the the solution $x_+$ is of BR-BR type.}
\label{n4112}
\end{figure}

\begin{center}
{\small{\it a.3. Four real roots}}
\end{center}

The situation is similar to the case a.2 and with two more roots, there is a
third admissible region only. We now have: $x_- \leq e_1 < e_2 \leq x_0 \leq e_3 < e_4 \leq x_+$,
depending on which of these intervals the line $x=1$ belongs to. The solutions $x_-$
(BB to BC) and $x_+$ (BR only), are described by equation (\ref{generic4}), and $x_0$ is obtained,
by adding a purely imaginary half-period $\omega_3$ to the argument. This last solution
is oscillating with a period $2\omega_1$, and avoids any singularity when
$0<e_2$ (Fig. \ref{n4113}).

\begin{figure}[h]
\includegraphics[angle=0,scale=.46]{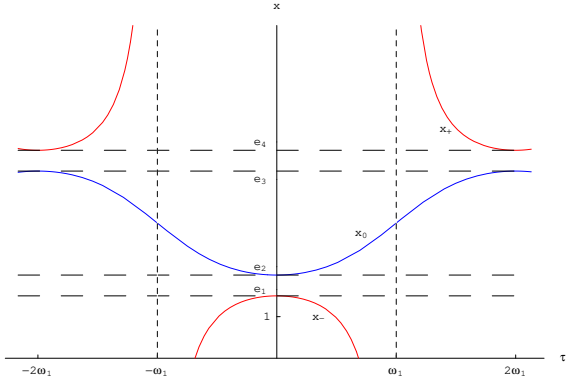}
\caption{Case B.1.a.3 -- four real roots. There are three solutions: $x_0$ (oscillating), and
$x_{\pm}$ (BR only, BB to BC) given by (\ref{generic4}) and (\ref{WWW}), respectively.}
\label{n4113}
\end{figure}

\subsubsection{$\sigma_6=0$ -- one double root}

\begin{center}
{\small{\it b.1. $\sigma_5<0$ -- two real and two complex roots}}
\end{center}

With the appearance of multiple roots, the solutions simplify significantly. Firstly,
it is possible to express the roots themselves in a suitably short form. Secondly,
the elliptic function $\wp$ reduces to a trigonometric or a hyperbolic function.

In this case, the double root is real, and the remaining two roots are complex.
Using equation (\ref{reduced4}), we can easily obtain a double root
\begin{equation}
    e_1 = \frac{-3q}{4p+2\sqrt{p^2+12r}}-\frac{a_3}{4a_4}, \label{double_root}
\end{equation}
which applied gives an appropriate solution
\begin{equation}
\label{WWW}
    x_{\pm} = \frac{6W_1''}
                {W_1'''+\sqrt{3W_1''W_1^{(4)}-W_1'''^2}
                \sinh\left[\sqrt{\frac12W_1''}(\tau-\tau_0)\right]},
\end{equation}
where the index indicates that the value $W(x)$ should be taken at $x=e_1$.
As in 1.a.2, this formula incorporates both $x_+ >e_1$ and $x_- <e_1$. Which of
these should be used depends on whether $e_1 <1$ or $1<e_1$, respectively. Also, there
is an unstable static solution $x_0=e_1$ and the solutions $x_{\pm}$
which approach it from either Big-Bang or Big-Rip (which differs
from standard case of an Einstein Static Universe - Fig.
\ref{n4121}).

\begin{figure}[h]
\includegraphics[angle=0,scale=.46]{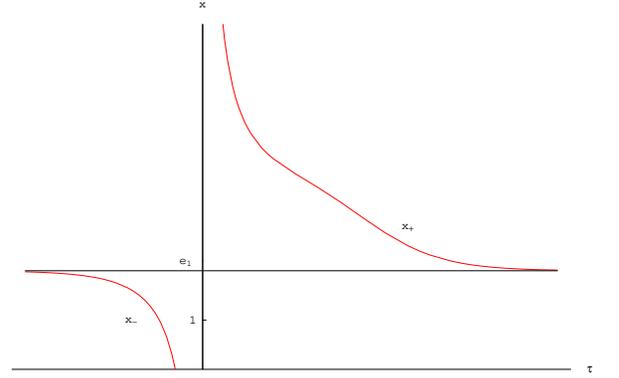}
\caption{Case B.1.b.1 -- two real and two complex roots. There is an unstable static solution
$x=e_1$ and the two asymptotic solutions $x_{\pm}$ given by (\ref{WWW}). One of them
approaches Big-Bang (MBB type) and another approaches Big-Rip (MBR type).}
\label{n4121}
\end{figure}

\begin{center}
{\small{\it b.2. $\sigma_5>0$ -- four real roots}}
\end{center}

Depending on the relative values of the roots, three situations are possible.
Denoting a double root by $e_1$, given by the same formula as before, we always
have a stable static solution $x_0=e_1$.  Since we know $e_1$, it is also possible
to obtain the remaining roots, and check which of the following is applicable.
If $e_2<e_1<e_3$,
\bea
\label{xpm1b2}
    &&x_{\pm} = e_2 +
    \\ && \frac{(e_2-e_1)(e_2-e_3)}
            {e_1-e_2+(e_1-e_3)\tan\left[\frac{1}{2}\sqrt{a_4(e_1-e_2)(e_3-e_1)}
            (\tau-\tau_0)\right]^2},\nonumber
\eea
with the choice of the + or - ``branch'' depending on whether $1<e_2$, or $e_3<1$
(Fig. \ref{n4122a}).

\begin{figure}[h]
\includegraphics[angle=0,scale=.46]{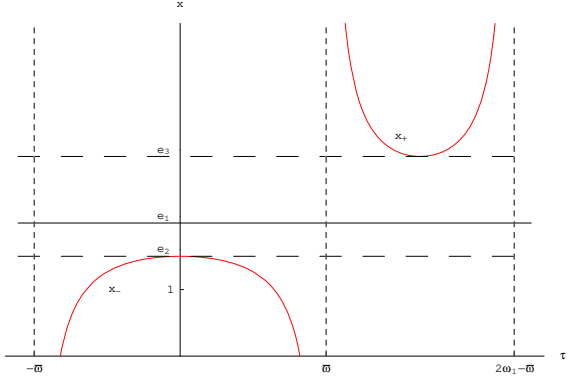}
\caption{Case B.1.b.2 -- the double root is between the other two
roots. There is a stable static solution $x=e_1$ and the two solutions given by
(\ref{xpm1b2}) (of BB-BC and BR-BR type).}
\label{n4122a}
\end{figure}

The second possibility is that $e_1<e_2<e_3$, where the solutions in the regions adjacent to
$e_1$, that is, $0\leq x < e_1$ and $e_1 < x \leq e_2$, tend to $e_1$
asymptotically, and $x_0=e_1$ becomes unstable.
The respective formulas are (Fig. \ref{n4122b})
\bea
    &&x_+ = e_2 +  \\
    && \frac{(e_2-e_1)(e_2-e_3)}
            {e_1-e_2+(e_3-e_1)\tanh\left[\frac{1}{2}\sqrt{a_4(e_2-e_1)(e_3-e_1)}(\tau-\tau_0)\right]^2},
            \nonumber \label{xp1b21}\\
    &&x_{-1,2} = e_2 + \\
    && \frac{(e_2-e_1)(e_2-e_3)}
        {e_1-e_2+(e_3-e_1)\coth\left[\frac{1}{2}\sqrt{a_4(e_2-e_1)(e_3-e_1)}(\tau-\tau_0)\right]^2}.
            \nonumber \label{x-12b21}
\eea
For a finite value of $\tau$, $x$ reaches infinity and
\bea
    \tau-\tau_0=\varpi=\frac{2}{\sqrt{a_4(e_1-e_2)(e_1-e3)}}\mathrm{artanh}
    \left(\sqrt{\frac{e_1-e_3}{e_1-e_2}}\right).\nonumber
\eea

\begin{figure}[h]
\includegraphics[angle=0,scale=.46]{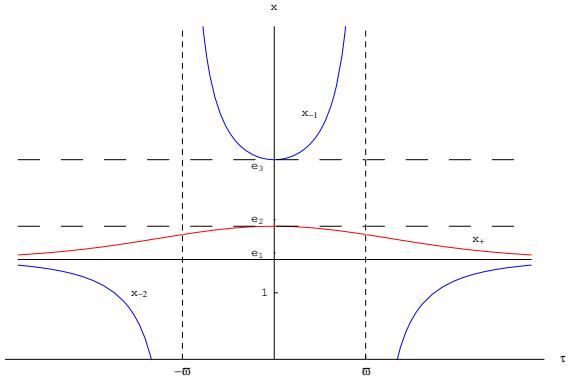}
\caption{Case B.1.b.2 -- the smallest root is double.
There is an unstable static (US) solution $x=e_1$, and the three other solutions $x_+$
(MUS) and $x_{-1,2}$ (MBB, BR-BR) given by (\ref{xp1b21}) and (\ref{x-12b21}), respectively.}
\label{n4122b}
\end{figure}

Lastly, when $e_2<e_3<e_1$, the situation is similar to the previous one,
only the asymptotic branch of $x_{-1}$ is now the upper one, with
\bea
    &&x_+ = e_2 + \\
    &&\frac{(e_2-e_1)(e_2-e_3)}
            {e_1-e_2+(e_3-e_1)\coth\left[\frac{1}{2}\sqrt{a_4(e_1-e_2)(e_1-e_3)}
            (\tau-\tau_0)\right]^2},\nonumber \label{xp1b22}\\
    &&x_{-1,2} = e_2 + \\
    &&\frac{(e_2-e_1)(e_2-e_3)}
            {e_1-e_2+(e_3-e_1)\tanh\left[\frac12\sqrt{a_4(e_1-e_2)(e_1-e_3)}
            (\tau-\tau_0)\right]^2},\nonumber \label{x-12b22}
\eea
and the static solution $x_0=e_1$ is also unstable (Fig. \ref{n4122c}).
Here we have
\bea
    \varpi=\frac{2}{\sqrt{a_4(e_1-e_2)(e_1-e_3)}}\mathrm{artanh}
    \left(\sqrt{\frac{e_1-e_2}{e_1-e_3}}\right).\nonumber
\eea

\begin{figure}[h]
\includegraphics[angle=0,scale=.46]{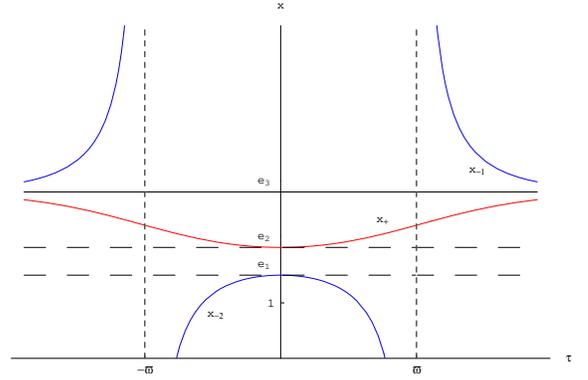}
\caption{Case B.1.b.2 -- the greatest root is double.
There is an unstable static solution $x=e_1$ and the other solutions
$x_+$ (oscillating) and $x_{-1,2}$ (from BB to BC, bounce to BR) given by (\ref{xp1b22}) and (\ref{x-12b22}),
respectively.}
\label{n4122c}
\end{figure}

\subsubsection{$\sigma_6=\sigma_5=0$, $\sigma_4\ne 0$ -- two double roots}

\begin{center}
{\small{\it c.1. $\sigma_1>0$ -- four real roots}}
\end{center}

The formula for the roots is further simplified to:
\begin{equation}
    e_{1,2} = -\frac{a_3}{4a_4}\mp\sqrt{-\frac{1}{2}p}. \label{2_double_real}
\end{equation}
Here $-2p$ is equal to the discriminant of the equation $(x-e_1)(x-e_2)$, and is always positive.
As for the solutions, we have two static, unstable solutions corresponding to $x_0=e_{1,2}$,
and the remaining two are given by (Fig. \ref{n4131})
\bea
    &&x_+ = e_1 + \nonumber \\
    &&\sqrt{-\frac{1}{2}p}\left\{1+\tanh\left[\sqrt{-\frac{1}{2}pa_4}(\tau-\tau_0)\right]
    \right\}, \label{xp1c1} \\
    &&x_{-1,2} = e_1 + \nonumber \\
    &&\sqrt{-\frac{1}{2}p}\left\{1+\coth\left[\sqrt{-\frac{1}{2}pa_4}(\tau-\tau_0)\right]\right\}.
    \label{x-121c1}
\eea

\begin{figure}[h]
\includegraphics[angle=0,scale=.46]{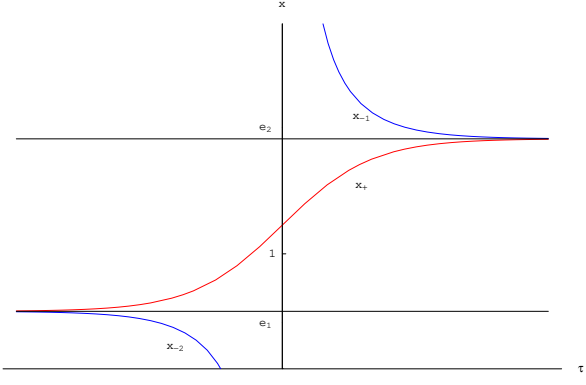}
\caption{Case B.1.c.1 -- four real roots, two of them are double. There are two
unstable static solutions $x_0=e_1, x_0=e_2$ and the three other
solutions $x_+$ and $x_{-1,2}$ given by (\ref{xp1c1}) and (\ref{x-121c1}), respectively.
The solution $x_+$ asymptotes from one unstable static solution to the other (MUS-MUS type)
while $x_{-1}$ is of MBR type and $x_{-2}$ is of MBB type.}
\label{n4131}
\end{figure}

\begin{center}
{\small{\it c.2. $\sigma_1<0$ -- four complex roots}}
\end{center}

Similarly to 1.a.1, the polynomial $W$ is always positive, but the solution is
simplified to
\begin{equation}
\label{x01c2}
    x_0 = -\frac{a_3}{4a_4}+\sqrt{\frac{1}{2}p}\,\tan\left[\sqrt{\frac{1}{2}pa_4}
    (\tau-\tau_0)\right],
\end{equation}
with $p$ always positive this time (Fig. \ref{n4132}).

\begin{figure}[h]
\includegraphics[angle=0,scale=.46]{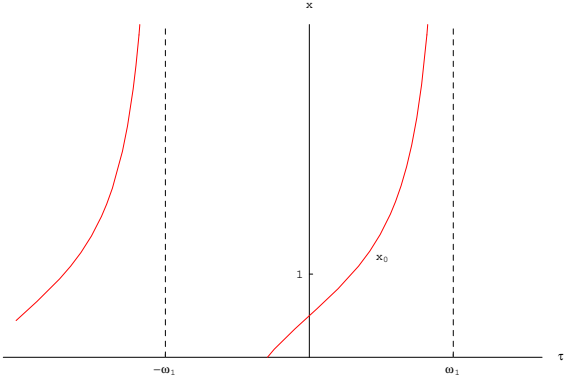}
\caption{Case B.1.c.2 -- four complex roots. The solution $x_0$ is given by
(\ref{x01c2}) and is of BB-BR type.}
\label{n4132}
\end{figure}

\subsubsection{$\sigma_6=\sigma_5=\sigma_4=0$, $\sigma_3\ne 0$ -- one triple root}

The multiple root now becomes:
\begin{equation}
    e_1 = -\frac{a_3}{4a_4}-\frac{1}{2}\sqrt[3]{q},
\end{equation}
and the fourth root is:
\begin{equation}
    e_2 = -\frac{a_3}{4a_4}+\frac{3}{2}\sqrt[3]{q}.
\end{equation}
There are two subcases, with $e_1<e_2$ and $e_1>e_2$, respectively, but the solution
for both ``branches'' in each subcase is given by the same formula:
\begin{equation}
\label{xpm1d}
    x_{\pm} = \frac{a_4e_1(e_1-e_2)^2(\tau-\tau_0)^2-4e_2}{a_4(e_1-e_2)^2(\tau-\tau_0)^2-4}.
\end{equation}
Infinite $x$ is reached for a finite value of $\tau$:
\bea
    \tau-\tau_0 = \varpi =
    \frac{2}{\sqrt{|a_4|}|e_1-e_2|}.\nonumber
\eea
There is also the unstable, static solution $x_0=e_1$ (Figs. \ref{n414a},\ref{n414b}).

\begin{figure}[h]
\includegraphics[angle=0,scale=.46]{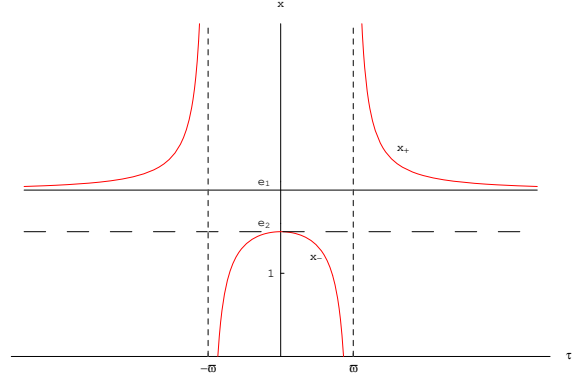}
\caption{Case B.1.d -- one triple root which is greater than a single root.
There is an unstable static solution $x_0=e_1$ and the other solutions $x_{\pm}$
(BB-BC and MBR type) are given by (\ref{xpm1d}).}
\label{n414a}
\end{figure}

\begin{figure}[h]
\includegraphics[angle=0,scale=.46]{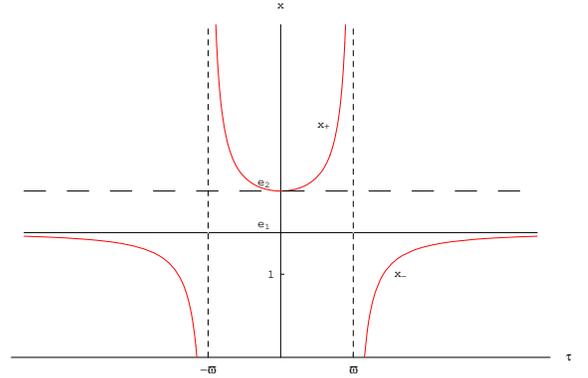}
\caption{Case B.1.d -- one triple root which is smaller than a single root.
There is an unstable static solution $x_0=e_1$ and the other two solutions
$x_{\pm}$ (BR to BR, BB only) are given by (\ref{xpm1d}).}
\label{n414b}
\end{figure}

\subsubsection{$\sigma_6=\sigma_5=\sigma_4=\sigma_1=0$ -- one quadruple root}

This is the simplest case, immediately integrable to:
\begin{equation}
\label{xpm1e}
    x_{\pm} = e_1 + \frac{1}{\sqrt{a_4}(\tau-\tau_0)},
\end{equation}
where the root is:
\begin{equation}
\label{e_11e}
    e_1 = -\frac{a_3}{4a_4}.
\end{equation}
Again, we have two ``branches'' separated by the static, unstable solution
$x_0=e_1$ (Fig. \ref{n415}).

\begin{figure}[h]
\includegraphics[angle=0,scale=.46]{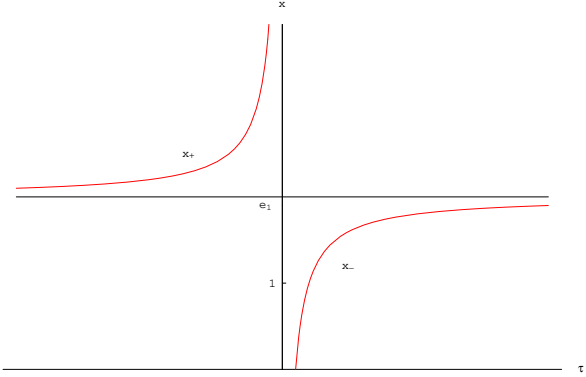}
\caption{Case B.1.e -- one quadrupole root.
There is an unstable static (US) solution and the solutions $x_{\pm}$
given by (\ref{xpm1e}. They are of MBR and MBB type.}
\label{n415}
\end{figure}

\subsection{$a_4<0$}

\subsubsection{$\sigma_6\ne 0$ -- simple roots}

\begin{center}
{\small{\it a.1. Four complex roots}}
\end{center}

This case is clearly impossible, as the polynomial $W$ is everywhere negative.

\begin{center}
{\small{\it a.2. Two complex and two real roots}}
\end{center}

There is only one possible solution, as $W(x)\geq 0$ when $e_1\geq x\geq e_2$,
where $e_1,e_2$ are the real roots. The general formula (\ref{generic4}) still holds, but
the behaviour of the solution is qualitatively different. It has a real period of
$2\omega_1$, defined as before, and avoids any singularity, provided that $e_1>0$
(Fig. \ref{n4212}).

\begin{figure}[h]
\includegraphics[angle=0,scale=.46]{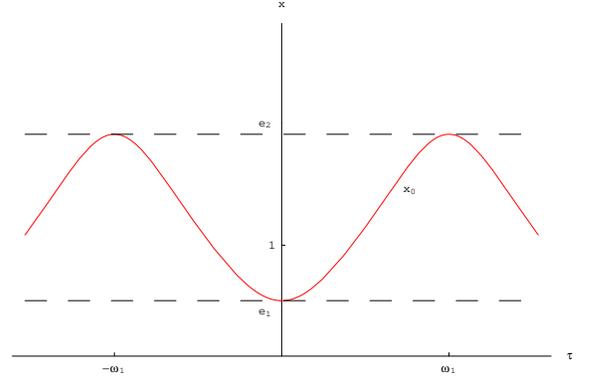}
\caption{Case B.2.a.2 -- two complex and two real roots. This is an oscillating
non-singular (ONS) solution $x_0$, given by the formula (\ref{generic4}).}
\label{n4212}
\end{figure}

\begin{center}
{\small{\it a.3. Four real roots}}
\end{center}

There are only two possible solutions when $e_1<1<e_2$ or $e_3<1<e_4$. They both are also
given by the general formula (\ref{generic4}), independently in each case. They are oscillating, and moreover,
and have the same period $2\omega_1$, so that, if $0<e_1$ no singularity is present at all
(Fig. \ref{n4213}).

\begin{figure}[h]
\includegraphics[angle=0,scale=.46]{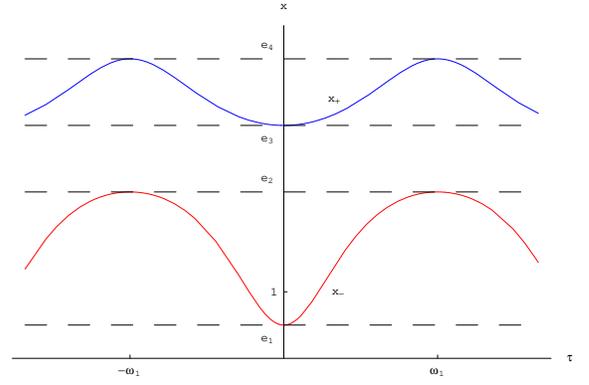}
\caption{Case B.2.a.3 -- four real roots. This is a ``double'' oscillating
case, which means we have the two oscillating solutions $x_{\pm}$ of the same period
$2\omega_1$ given by the formula (\ref{WWW}). They are non-singular provided $e_1>0$.}
\label{n4213}
\end{figure}

\subsubsection{$\sigma_6=0$ -- one double root}

\begin{center}
{\small{\it b.1. $\sigma_5<0$ -- two real and two complex roots}}
\end{center}

There is only one possible solution: $x_0=e_1$, with the root given by (\ref{double_root}).
As $W$ is negative everywhere else, this solution is stable.

\begin{center}
{\small{\it b.2. $\sigma_5>0$ -- four real roots}}
\end{center}

Here, there are two subcases, both with the static solution $x_0=e_1$. The first one
occurs when $e_2<e_1<e_3$, and there are two other solutions asymptotic to $x_0$
(Fig. \ref{n4222a})
\bea
    &&x_+ = e_2 +  \label{x2b21p}\\
    &&\frac{(e_2-e_1)(e_2-e_3)}
            {e_1-e_2+(e_3-e_1)\coth\left[\frac{1}{2}\sqrt{|a_4|(e_1-e_2)(e_3-e_1)}
            (\tau-\tau_0)\right]^2},\nonumber
\eea
\bea
    &&x_- = e_2 +  \label{x2b21m}\\
    && \frac{(e_2-e_1)(e_2-e_3)}
            {e_1-e_2+(e_3-e_1)\tanh\left[\frac{1}{2}\sqrt{|a_4|(e_1-e_2)(e_3-e_1)}
            (\tau-\tau_0)\right]^2}.\nonumber
\eea

If $e_1<e_2$ or $e_3<e_1$, $x_0$ becomes stable, and there is only one other oscillating
solution between $e_2$ and $e_3$ (Fig. \ref{n4222b})
\bea
    &&x_+ = e_2 +  \label{xp2b22}\\
    &&\frac{(e_2-e_1)(e_2-e_3)}
            {e_1-e_2+(e_1-e_3)\tan\left[\frac{1}{2}\sqrt{|a_4|(e_1-e_2)(e_1-e_3)}
            (\tau-\tau_0)\right]^2}.\nonumber
\eea

\begin{figure}[h]
\includegraphics[angle=0,scale=.46]{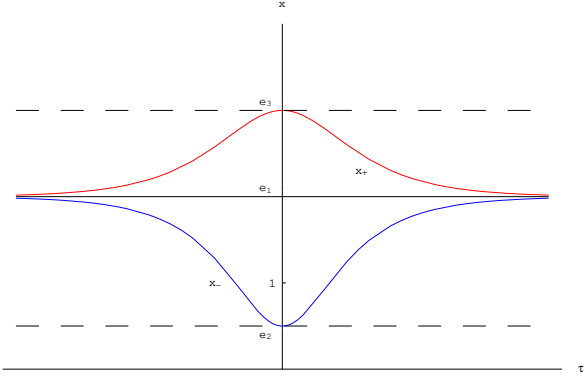}
\caption{Case B.2.b.2 -- one double root which is between the two single roots.
There is an unstable static (US) solution
$x_0=e_1$ and the two asymptotic solutions $x_{\pm}$ given by (\ref{x2b21p})
and (\ref{x2b21m}).}
\label{n4222a}
\end{figure}

\begin{figure}[h]
\includegraphics[angle=0,scale=.46]{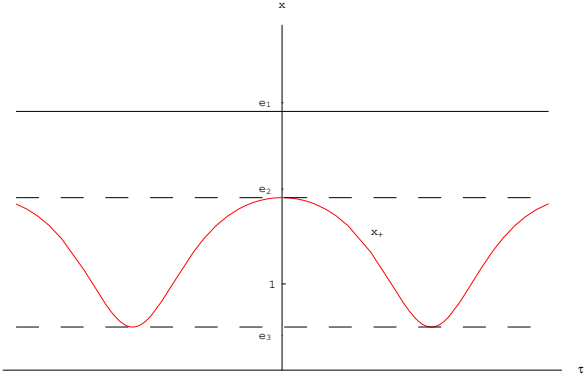}
\caption{Case B.2.b.2 -- one double root greater than the two single roots.
There is a stable static (SS) solution
$x_0=e_1$ and an oscillating solution $x_+$ (ONS type) given by (\ref{xp2b22}).}
\label{n4222b}
\end{figure}

\subsubsection{$\sigma_6=\sigma_5=0$, $\sigma_4\ne 0$ -- two double roots}

\begin{center}
{\small{\it c.1. $\sigma_1>0$ -- four real roots}}
\end{center}

The only possible solutions here, are two static, stable ones: $x_0=e_{1,2}$.
The roots are given by formula (\ref{2_double_real}).

\begin{center}
{\small{\it c.2. $\sigma_1<0$ -- four complex roots}}
\end{center}

\noindent $W$ is negative for all $x$, so this case is impossible.

\subsubsection{$\sigma_6=\sigma_5=\sigma_4=0$, $\sigma_3\ne 0$ -- one triple root}

Essentially this is just one case, but depending on the position of the triple root,
the non-static solution
\begin{equation}
\label{xpm2d1}
    x_{\pm} = \frac{4e_2-a_4e_1(e_1-e_2)^2(\tau-\tau_0)^2}{4-a_4(e_1-e_2)^2(\tau-\tau_0)^2},
\end{equation}
has either a maximum, for $e_1<e_2$, or a minimum, for $e_1>e_2$. Also,
$x_0=e_1$ is an unstable static solution (Figs. \ref{n424a},\ref{n424b}).

\begin{figure}[h]
\includegraphics[angle=0,scale=.46]{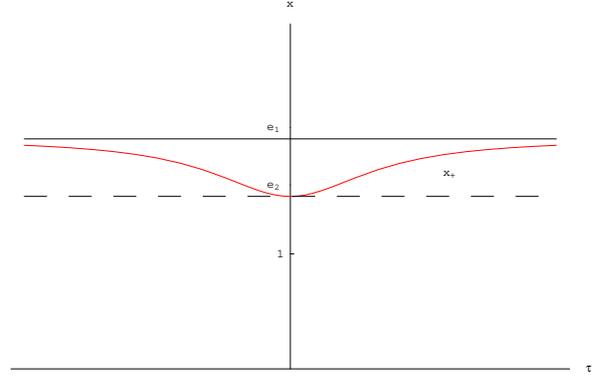}
\caption{Case B.2.d -- one triple root greater than a single root. There is an unstable
static (US) solution
$x_0=e_1$ and a monotonic solution $x_+$ given by (\ref{xpm2d1}) which has a minimum.}
\label{n424a}
\end{figure}

\begin{figure}[h]
\includegraphics[angle=0,scale=.46]{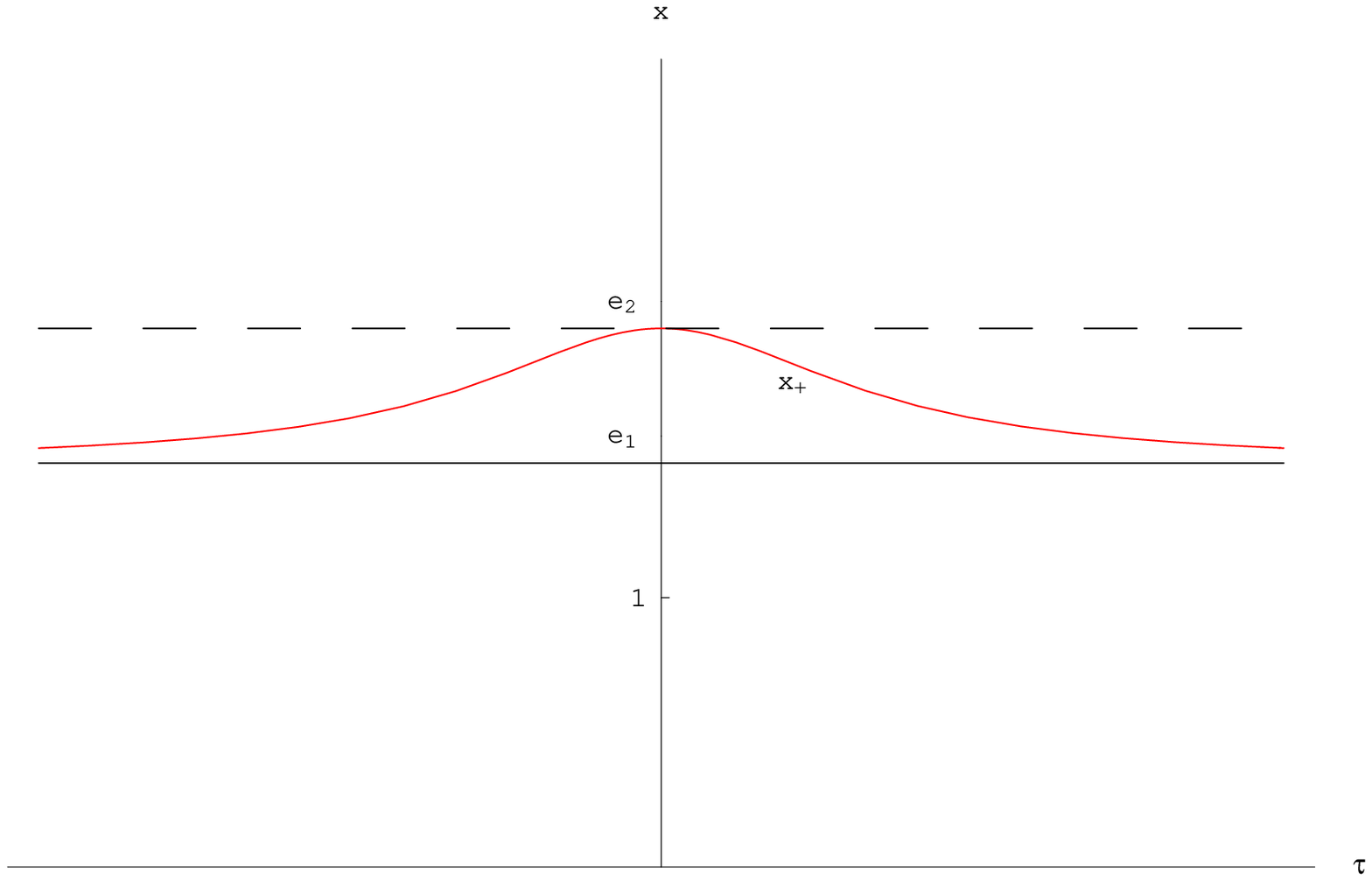}
\caption{Case B.2.d -- one triple root smaller that a single root. There is an unstable static solution
$x_0=e_1$ and a monotonic solution $x_+$ given by (\ref{xpm2d1}) which has a maximum.}
\label{n424b}
\end{figure}

\subsubsection{$\sigma_6=\sigma_5=\sigma_4=\sigma_1=0$ -- one quadruple root}

There is only one point where $W$ is not negative, thus yielding just the
stable, static solution of $x_0=e_1$, where $e_1$ is given by the formula
(\ref{e_11e}).

\subsection{$a_4 = 0$}

This case is essentially reduced to a cubic polynomial case
considered in the Appendix A.


\begin{thebibliography}{99}

\bibitem{striwall} A. Vilenkin, Phys. Rep. {\bf 121} (1985), 265;
T. Vachaspati, and A. Vilenkin, Phys. Rev. D{\bf 35} (1987), 1131.

\bibitem{AJIII} M.P. D\c{a}browski, and J. Stelmach, Astron. Journ. {\bf 97}
(1989), 978 (astro-ph/0410334).

\bibitem{perlmutter} S. Perlmutter {\it et al.}, Astroph. J. {\bf 517},
(1999) 565; A. G. Riess {\it et al.} Astron. J. {\bf 116}
(1998) 1009; A.G. Riess {\it et al.}, Astroph. J. {\bf 560} (2001), 49;
{\it ibidem} {\bf 594} (2003), 1.

\bibitem{supernovae} J.L. Tonry {\it et al.}, Astroph. J. {\bf 594} (2003), 1;
M. Tegmark {\it et al.}, Phys. Rev. D{\bf 69} (2004), 103501.

\bibitem{caldwell99} R.R. Caldwell, Phys. Lett. B{\bf 545} (2002), 23 (astro-ph/9908168).

\bibitem{instabilities} S.M. Carroll, M. Hofman, and M. Trodden,
Phys. Rev. D{\bf 68} (2003), 023509; S.D.H. Hsu, A. Jenkins, and
M.B. Wise, Phys. Lett. B{\bf 597} (2004), 270.

\bibitem{nojiri04} S. Nojiri, and S.D. Odintsov, Phys. Lett. B{\bf 595} 
(2004), 1; Phys. Rev. D{\bf 70} (2004), 103522.

\bibitem{he} S.W. Hawking, and G.F.R. Ellis, {\it The
    Large-scale Structure of Space-time} (Cambridge Univ. Press, 1999).

\bibitem{visser} M. Visser, {\it Lorentzian Wormholes} (Springer, New York, 1996).

\bibitem{kaloper} C. Csaki, N. Kaloper, and J. Terning, Ann. Phys. (N.Y.) 
{\bf 317} (2005), 410. 

\bibitem{barrow04} J.D. Barrow, Class. Quantum Grav. {\bf 21}, L79
(2004); {\it ibidem} {\bf 21}, 5619 (2004), J.D. Barrow, and Ch. Tsagas,
Class. Quantum Grav. {\bf 22} (2005), 1563; L. Fernandez-Jambrina, and R. Lazkoz, Phys. Rev.
D{\bf 70}, 121503 (2004).

\bibitem{braneIa} M.P. D\c{a}browski, W. God{\l }owski, and M. Szyd{\l }owski,
Intern. Journ. Mod. Phys. D {\bf 13} (2004), 1669.

\bibitem{john88} J.D. Barrow, Nucl. Phys. B{\bf 310} (1988), 743.

\bibitem{Pollock88} M.D. Pollock, Phys. Lett. B{\bf 215} (1988),
635.

\bibitem{starob00} B. Boisseau, G. Esposito-Far\'ese, D. Polarski,
and A.A. Starobinsky, Phys. Rev. Lett. {\bf 85} (2000), 2236; A.A.
Starobinsky, Grav. Cosmol. {\bf 6} (2000), 157.

\bibitem{kiritsis} A. Kehagias, and E. Kiritsis, JHEP 9911 (1999), 022.

\bibitem{chiba} T. Chiba, T. Okabe, and M. Yamaguchi, Phys. Rev. D{\bf 62} (2000), 023511.

\bibitem{Hann02} S. Hannestad, and E. M\"orstell, Phys. Rev. D {\bf 66}, 063508 (2002)

\bibitem{Frampt02} P.H. Frampton, Phys. Lett. B{\bf 557} (2003), 135.

\bibitem{Frampt03} P.H. Frampton, Phys. Lett. B {\bf 555} (2003),
139.

\bibitem{FramTaka02} P.H. Frampton, and T. Takahashi, Phys. Lett. B{\bf 557} (2003), 135.

\bibitem{McInnes} B. McInnes, JHEP 0208 (2002), 029.

\bibitem{trodden1} S.M. Carroll, M. Hoffman, and M. Trodden, Phys. Rev. D {\bf 68} (2003),
                   023509.

\bibitem{trodden2} A. Melchiorri, L. Mersini, C.J. Odman, and M. Trodden, Phys. Rev. D {\bf 68}
                   (2003) 043509.

\bibitem{Bastero01} L. Mersini, M. Bastero-Gil and P. Kanti, Phys.
Rev. D{\bf 64} (2001), 043508.

\bibitem{Bastero02} M. Bastero-Gil, P.H. Frampton and L. Mersini,
Phys. Rev. D{\bf 65} (2002), 106002.

\bibitem{abdalla04} M. Abdalla, S. Nojiri, and S. Odintsov, Class. Quantum Grav. {\bf 22} (2005), L35.

\bibitem{Erickson} J.K. Erickson, R.R. Caldwell, P.J. Steinhardt, C.
Armendariz-Picon, and V. Mukhanov, Phys. Rev. Lett. {\bf 88} (2002), 121301.

\bibitem{BIphantom} J. Hao, and X. Li, Phys. Rev. D {\bf 68} (2003), 043501.

\bibitem{LiHao03} X. Li, and J. Hao, Phys. Rev. D{\bf 69} (2004) 107303.

\bibitem{singh03} P. Singh, M. Sami, and N. Dadhich, Phys. Rev. D {\bf 68} (2003),
                  043501.
\bibitem{nojiri031} S. Nojiri, and D. Odintsov, Phys. Lett. B{\bf 562} (2003), 147.

\bibitem{nojiri032} S. Nojiri, and D. Odintsov, Phys. Lett. B{\bf 565} (2003), 1.

\bibitem{nojiri033} S. Nojiri, and D. Odintsov, Phys. Lett. B{\bf 571} (2003), 1.

\bibitem{nojiri034} S. Nojiri, and S.D. Odintsov, Phys. Lett. B{\bf 576} (2003), 5.

\bibitem{simphan} M. Szyd{\l}owski, W. Czaja, and A. Krawiec, Phys. Rev. E{/bf 72} (2005), 036221.

\bibitem{onemli} V.K. Onemli, and R.P. Woodard, Class. Quantum
Grav. {\bf 19} (2002), 4607; Phys. Rev. D {\bf 70} (2004), 107301;
T. Brunier, V.K. Onemli, and R.P. Woodard, Class. Quantum Grav. {\bf 22} (2005),
59.

\bibitem{diaz} P.F. Gonzalez-Diaz, Phys. Rev. D {\bf 68} (2003),
021303; {\it ibid} D {\bf 69} (2004), 063522; Phys. Lett. B {\bf
586} (2004), 1; e-print: hep-th/0411070, P.F. Gonzalez-Diaz, and C.L. Sig\"{u}enza,
Nucl. Phys. B {\bf 697} (2004), 363.

\bibitem{zhang} B.B. Feng, X-L. Wang, and X-M. Zhang, Phys. Lett. B{\bf 607} (2005), 35;
Z-K. Guo, Y-S. Piao, and Y-Z. Zhang, Phys. Lett. B{\bf 594}
(2004), 247; B.B. Feng, M. Li, Y-S.Piao, and X-M. Zhang, e-print: astro-ph/0407432,
Z-K. Guo, Y-S. Piao, X-M. Zhang, and Y-Z. Zhang, Phys. Lett. B{\bf 608} (2005), 177;
Z-K. Guo, and Y-Z. Zhang, Phys. Rev. D{\bf 71} (2005), 023501.

\bibitem{cai} R-G. Cai and A. Wang, JCAP 0503 (2005), 002.

\bibitem{calcagni} G. Calcagni, Phys. Rev. D{\bf 71} (2005),
023511; gr-qc/0410111.

\bibitem{dabrowski93} M.P. D\c{a}browski, Journ. Math. Phys. {\bf 34}, 1447
(1993).

\bibitem{dabrowski95} M.P. D\c{a}browski, Astroph. J. {\bf 447}, 43
(1995).

\bibitem{inhosfs} M.P. D\c{a}browski, Phys. Rev. D {\bf 71} (2005), 103505.


\bibitem{phantom1} M.P. D\c{a}browski, T. Stachowiak and M. Szyd{\l }owski, Phys. Rev. D {\bf 68},
103519 (2003).

\bibitem{meissner91} K.A. Meissner and G. Veneziano, Phys. Lett.
B{\bf 267} (1991), 33; Mod. Phys. Lett. A{\bf 6} (1991), 1721.

\bibitem{superjim} J.E. Lidsey, D.W. Wands, and E.J. Copeland,
         Phys. Rep. {\bf 337} (2000), 343.

\bibitem{lazkoz1} L.P. Chimento Phys. Rev. D{\bf 65} (2002), 063517.

\bibitem{lazkoz2} J.M. Aguirregabiria, L.P. Chimento, A.S. Jacubi, and R. Lazkoz, Phys. Rev. D{\bf 67}
(2003), 083518.

\bibitem{ekpyrotic} J. Khoury {\it et al}, Phys. Rev. D {\bf 64}
(2001), 123522; P.J. Steinhardt, and N. Turok, Phys. Rev. D {\bf
65} (2002), 126003; J. Khoury, P.J. Steinhardt, and N. Turok, Phys.
Rev. Lett. {\bf 92} (2004), 031302.

\bibitem{triality} J.E. Lidsey, Phys. Rev. D {\bf 70} (2004), 041302.

\bibitem{jerk} M. Visser, Class. Quantum Grav. {\bf 21} (2004), 2603; U. Alam, V. Sahni,
               T.D. Saini, and A.A. Starobinsky, Mon. Not. R.
               Astron. Soc. 344 (2003), 1057.

\bibitem{snap} R.R. Caldwell, and M. Kamionkowski, J. Cosmol. Astropart. 
Phys. 0409 (2004), 009.

\bibitem{weinberg} S. Weinberg, {\it Gravitation and Cosmology}
(Wiley, New York, 1972).

\bibitem{FRWelliptic} R. Coquereaux, and A. Grossmann Ann. Phys. (N.Y.) {\bf 143} (1982),
296; M.P. D\c{a}browski, and J. Stelmach, Ann. Phys (N. Y.) {\bf 166} (1986),
422; M.P. D\c{a}browski, Ann. Phys (N. Y.) {\bf 248} (1996) 199;
Coquereaux and Grossmann, astro-ph/0101369.

\bibitem{abramovitz} M.Abramovitz, and I.A. Stegun, {\it Handbook
of Mathematical Functions} (Dover, New York, 1964).

\bibitem{lake} K. Lake, Class. Quantum Grav. {\bf 21} (2004),
L129.

\bibitem{krsachs66} J. Kristian, and R.K. Sachs, Astrophys. J.
{\bf 143} (1966), 379.

\bibitem{ellis70} G.F.R. Ellis, and M.A.H. MacCallum, Commun.
Math. Phys. {\bf 19} (1970), 31.






\end{thebibliography}
\end{document}